\documentclass[PRD,preprintnumbers,floatfix,aps,notitlepage,nofootinbib,twocolumn,showpacs,amssymb]{revtex4-1}
\usepackage{amsmath,amsfonts,bm}
\usepackage{graphicx}
\usepackage[colorlinks, linkcolor={red},citecolor={blue}]{hyperref}

\begin{document}
\title{Hairy black holes by gravitational decoupling}
\author{J. Ovalle}
\email[]{jorge.ovalle@physics.slu.cz}
\affiliation{Research Centre of Theoretical Physics and Astrophysics,
Institute of Physics, Silesian University in Opava, CZ-746 01 Opava, Czech Republic.
}
\author{R. Casadio}
\email{casadio@bo.infn.it}
\affiliation{Dipartimento di Fisica e Astronomia,
Alma Mater Universit\`a di Bologna,
40126 Bologna, Italy
\\
Istituto Nazionale di Fisica Nucleare, 
Sezione di Bologna, 40127 Bologna, Italy}
\author{E. Contreras}
\email{econtreras@usfq.edu.ec}
\affiliation{Departamento de F\'isica, Colegio de Ciencias e Ingenier\'ia,
Universidad San Francisco de Quito, Quito, Ecuador.
}
\author{A.~Sotomayor}
\email{adrian.sotomayor@uantof.cl}
\affiliation{Departamento de Matem\'aticas, Universidad de Antofagasta, Antofagasta, Chile.}
\begin{abstract}
Black holes with hair represented by generic fields surrounding the central source of the vacuum Schwarzschild
metric are examined under the minimal set of requirements consisting of
i) the existence of a well defined event horizon and ii) the strong or dominant energy condition for the hair outside
the horizon.
We develop our analysis by means of the gravitational decoupling approach.
We find that trivial deformations of the seed Schwarzschild vacuum preserve the energy conditions
and provide a new mechanism to evade the no-hair theorem based on a primary hair associated with
the charge generating these transformations.
Under the above conditions i) and ii), this charge consistently increases the entropy from the minimum value
given by the Schwarzschild geometry.
As a direct application, we find a non-trivial extension of the Reissner-Nordstr\"om black hole
showing a surprisingly simple horizon.
Finally, the non-linear electrodynamics generating this new solution is fully specified.
\end{abstract} 
\maketitle


\section{Introduction}
Black holes are among the most extraordinary objects in the Universe.
The direct observations of black holes through the detection of gravitational
waves~\cite{Abbott:2016blz,Abbott:2017oio} and the reconstruction of the black hole
shadow~\cite{Akiyama:2019cqa} have definitely raised them from being exotic solutions
of general relativity, to the status of real astrophysical systems with well determined
characteristics.
\par
The original {\em no-hair\/} conjecture states that black hole solutions should not carry any other
charges~\cite{Ruffini:1971bza} except three fundamental parameters, namely the mass $M$,
angular momentum $J$ and electric charge $Q$~\cite{Hawking:1971vc}.
However, there could exist other charges associated with inner gauge symmetries
(and fields), and it is now known that black holes could also have (soft) quantum
hair~\cite{Hawking:2016msc}.
The {\em general\/} existence of {\em hairy\/} black hole solutions is precisely the topic under
study in this article.
\par
For a long time different scenarios have been studied for circumventing the no-hair theorem
(see Refs.~\cite{Sotiriou:2011dz,Babichev:2013cya,Cisterna:2014nua,Sotiriou:2013qea,
Antoniou:2017acq,Antoniou:2017hxj,Grumiller:2019fmp}
for some recent works and Refs.~\cite{Volkov:1989fi,Kanti:1995vq,Kanti:1997br,Zloshchastiev:2004ny}
for earlier works).
For instance, scalar fields have played a preponderant role, mainly due to their simplicity
and also in analogy with particle physics and cosmology (see also Refs.~\cite{Martinez:2004nb,Herdeiro:2015waa,Sotiriou:2015pka} and references therein).
In this paper, following our previous work~\cite{Ovalle:2018umz}, instead of considering specific
fundamental fields to generate hair in black holes, we shall just assume the presence of a
generic source in addition to the one generating the vacuum Schwarzschild geometry.
We then impose a minimal set of conditions we expect should hold for hairy black holes,
namely: 
i) we require that the system has a well-defined event horizon and
ii) the additional source is described by a conserved energy-momentum tensor $\theta_{\mu\nu}$
which satisfies either the strong (SEC) or the dominant energy condition (DEC) in the region outside
the event horizon.
\par
From the technical point of view, we will also assume the energy-momentum tensor $\theta_{\mu\nu}$
is {\em decoupled from the vacuum\/} by means of the extended gravitational decoupling
method (EGD henceforth) of Ref.~\cite{Ovalle:2019qyi}.
This approach was originally introduced in Ref.~\cite{Ovalle:2017fgl} in the form of the so-called
minimal geometric deformation (MGD)~\cite{Ovalle:2007bn,Ovalle:2020fuo} 
(for some earlier works on the MGD, see for instance Refs.~\cite{Ovalle:2008se,Ovalle:2010zc,
Casadio:2012pu,Casadio:2012rf,Ovalle:2013vna,Casadio:2013uma,Ovalle:2013xla,
Ovalle:2014uwa,Casadio:2015jva,Casadio:2015gea,Ovalle:2015nfa,Cavalcanti:2016mbe},
and Refs.~\cite{daRocha:2017cxu,daRocha:2017lqj,Fernandes-Silva:2017nec,Casadio:2017sze,
Fernandes-Silva:2018abr,Contreras:2018vph,Contreras:2018gzd,Contreras:2018nfg,
Panotopoulos:2018law,daRocha:2019pla,Heras:2019ibr,Rincon:2019jal,
daRocha:2020rda,Contreras:2020fcj,Arias:2020hwz,daRocha:2020jdj,Tello-Ortiz:2020euy,
daRocha:2020gee,Meert:2020sqv} for some recent applications).
Analogously to the electro-vacuum and scalar-vacuum cases, in this work we will thus consider
a Schwarzschild black hole surrounded by a spherically symmetric ``tensor-vacuum'',
represented by the aforementioned $\theta_{\mu\nu}$.
Following the EGD, we can separate the complete Einstein field equations and
obtain the ``quasi-Einstein" equations for $\theta_{\mu\nu}$
[See Eqs.~\eqref{ec1d}-\eqref{ec3d} below].
These are precisely the equations of motion for the deformed Schwarzschild vacuum,
which after decoupling, contains five unknown functions that can be further analysed
by imposing the two conditions discussed above.
\par
The paper is organised as follows:
in Section~\ref{Sgd}, we first review the fundamentals of the EGD approach 
to a spherically symmetric system containing two sources;
in Section~\ref{Shbh}, after imposing a simple condition to guarantee a well defined horizon,
and after considering the strong and dominant energy conditions on $\theta_{\mu\nu}$,
we show a new way to evade the no-hair theorem.
We apply this to generate two new families of hairy black holes containing primary hairs.
As a special case, we show a non-trivial extension
of the Reissner-Nordstr\"om  black hole associated with a non-linear
electrodynamics; finally, we summarize our conclusions in Section~\ref{Scon}.
\section{Gravitational Decoupling}
\label{Sgd}
In this Section, we briefly review the EGD for spherically symmetric gravitational systems
described in detail in Ref.~\cite{Ovalle:2019qyi}. The gravitational decoupling approach and its extended version EGD are particularly attractive for at least three reasons~\cite{Ovalle:2017wqi,
Gabbanelli:2018bhs,Heras:2018cpz,Estrada:2018zbh,Sharif:2018toc,Morales:2018nmq,
Sharif:2018pzr,Morales:2018urp,Estrada:2018vrl,Sharif:2018tiz,Ovalle:2018ans,
Contreras:2019fbk,Maurya:2019wsk,Contreras:2019iwm,Contreras:2019mhf,
Gabbanelli:2019txr,Estrada:2019aeh,Ovalle:2019lbs,Ovalle:2019lbs,Maurya:2019hds,Hensh:2019rtb,Cedeno:2019qkf,Leon:2019abq,Torres:2019mee,
Casadio:2019usg,Singh:2019ktp,Maurya:2019noq,Sharif:2019mjn,Singh:2019ktp,
Abellan:2020wjw,Sharif:2020vvk,Tello-Ortiz:2020ydf,
Maurya:2020rny,Rincon:2020izv,Sharif:2020arn,Maurya:2020gjw}:
a) it allows for extending known (seed) solutions of the Einstein field equations into 
more complex ones;
b) it can be used to systematically reduce (decouple) a complex energy-momentum
tensor $T_{\mu\nu}$ into simpler components; and
c) it can be used to find solutions in gravitational theories beyond Einstein's.
In light of the above, it is natural to apply the EGD for the purpose of describing
hairy black holes.
\par
Let us consider the Einstein field equations~\footnote{We use units with $c=1$ and $k^2=8\,\pi\,G_{\rm N}$,
where $G_{\rm N}$ is Newton's constant.}
\begin{equation}
\label{corr2}
G_{\mu\nu}
\equiv
R_{\mu\nu}-\frac{1}{2}\,R\, g_{\mu\nu}
=
k^2\,\tilde{T}_{\mu\nu}
\ ,
\end{equation}
with a total energy-momentum tensor containing two contributions,
\begin{equation}
\label{emt}
\tilde{T}_{\mu\nu}
=
T^{\rm}_{\mu\nu}
+
\theta_{\mu\nu}
\ ,
\end{equation}
where $T_{\mu\nu}$ is usually associated with some known solution of general relativity, 
whereas $\theta_{\mu\nu}$ may contain new fields or a new gravitational sector.
Since the Einstein tensor $G_{\mu\nu}$ satisfies the Bianchi identity, the total source must be
covariantly conserved,
\begin{equation}
\nabla_\mu\,\tilde{T}^{\mu\nu}=0
\ .
\label{dT0}
\end{equation} 
For spherically symmetric and static systems, the metric $g_{\mu\nu}$ can be written as
\begin{equation}
ds^{2}
=
e^{\nu (r)}\,dt^{2}-e^{\lambda (r)}\,dr^{2}
-r^{2}\,d\Omega^2
\ ,
\label{metric}
\end{equation}
where $\nu =\nu (r)$ and $\lambda =\lambda (r)$ are functions of the areal
radius $r$ only and $d\Omega^2=d\theta^{2}+\sin ^{2}\theta \,d\phi ^{2}$.
The Einstein equations~(\ref{corr2}) then read
\begin{eqnarray}
\label{ec1}
k^2\!
\left(
T_0^{\ 0}+\theta_0^{\ 0}
\right)
&=&
\frac 1{r^2}
-
e^{-\lambda }\left( \frac1{r^2}-\frac{\lambda'}r\right)
\\
\label{ec2}
k^2\!
\left(T_1^{\ 1}+\theta_1^{\ 1}\right)
&=&
\frac 1{r^2}
-
e^{-\lambda }\left( \frac 1{r^2}+\frac{\nu'}r\right)
\\
\label{ec3}
k^2\!
\left(T_2^{\ 2}+\theta_2^{\ 2}\right)
&=&
-\frac {e^{-\lambda }}{4}
\left(2\nu''+\nu'^2-\lambda'\nu'
+2\,\frac{\nu'-\lambda'}r\right)
\ ,
\end{eqnarray}
where $f'\equiv \partial_r f$ and $\tilde{T}_3^{{\ 3}}=\tilde{T}_2^{\ 2}$ due to the spherical symmetry.
By simple inspection, we can identify in Eqs.~\eqref{ec1}-\eqref{ec3} an effective density  
\begin{equation}
\tilde{\rho}
=
T_0^{\ 0}
+
\theta_0^{\ 0}
\ ,
\label{efecden}
\end{equation}
an effective radial pressure
\begin{equation}
\tilde{p}_{r}
=
-T_1^{\ 1}
-\theta_1^{\ 1}
\ ,
\label{efecprera}
\end{equation}
and an effective tangential pressure
\begin{equation}
\tilde{p}_{t}
=
-T_2^{\ 2}
-\theta_2^{\ 2}
\ .
\label{efecpretan}
\end{equation} 
Moreover, the anisotropy 
\begin{equation}
\label{anisotropy}
\Pi
\equiv
\tilde{p}_{t}-\tilde{p}_{r}
\end{equation}
usually does not vanish and the system of Eqs.~(\ref{ec1})-(\ref{ec3}) may be
treated as an anisotropic fluid~\cite{Herrera:1997plx,Mak:2001eb}.
\par
We next consider a solution to the Eqs.~\eqref{corr2} for the seed source $T_{\mu\nu}$
alone [that is, $\theta_{\mu\nu}=0$], which we write as
\begin{equation}
ds^{2}
=
e^{\xi (r)}\,dt^{2}
-e^{\mu (r)}\,dr^{2}
-
r^{2}\,d\Omega^2
\ ,
\label{pfmetric}
\end{equation}
where 
\begin{equation}
\label{standardGR}
e^{-\mu(r)}
\equiv
1-\frac{k^2}{r}\int_0^r x^2\,T_0^{\, 0}(x)\, dx
=
1-\frac{2\,m(r)}{r}
\end{equation}
is the standard general relativity expression containing the Misner-Sharp mass function $m=m(r)$.
The addition of the source $\theta_{\mu\nu}$ can then be accounted for by the extended
geometric deformation (EGD) of the seed metric~\eqref{pfmetric}, namely
\begin{eqnarray}
\label{gd1}
\xi 
&\rightarrow &
\nu\,=\,\xi+\alpha\,g
\\
\label{gd2}
e^{-\mu} 
&\rightarrow &
e^{-\lambda}=e^{-\mu}+\alpha\,f
\ , 
\end{eqnarray}
where $f$ and $g$ are respectively the geometric deformations for the radial and temporal metric
components, with the parameter $\alpha$ introduced to keep track of these deformations.
By means of Eqs.~(\ref{gd1}) and (\ref{gd2}), the Einstein equations~(\ref{ec1})-(\ref{ec3})
are separated in two sets:
A) one is given by the standard Einstein field equations with the energy-momentum tensor $T_{\mu\nu}$,
that is
\begin{eqnarray}
\label{ec1pf}
&&
k^2\,T_0^{\, 0}
=\frac 1{r^2}
-
e^{-\mu }\left( \frac1{r^2}-\frac{\mu'}r\right)\ ,
\\
&&
\label{ec2pf}
k^2
\,T_1^{\, 1}
=
\frac 1{r^2}
-
e^{-\mu}\left( \frac 1{r^2}+\frac{\xi'}r\right)\ ,
\\
&&
\label{ec3pf}
k^2
\strut\displaystyle
\,T_2^{\, 2}
=
-\frac {e^{-\mu }}{4}
\left(2\xi''+\xi'^2-\mu'\xi'
+2\,\frac{\xi'-\mu'}r\right)
\ ,
\end{eqnarray}
which is assumed to be solved by the seed metric~(\ref{pfmetric});
B) the second set contains the source $\theta_{\mu\nu}$ and reads
\begin{eqnarray}
\label{ec1d}
k^2\,\theta_0^{\ 0}
&=&
-\alpha\,\frac{f}{r^2}
-\alpha\,\frac{f'}{r}\ ,
\\
\label{ec2d}
k^2\,\theta_1^{\ 1}
+\alpha\,Z_1
&=&
-\alpha\,f\left(\frac{1}{r^2}+\frac{\nu'}{r}\right)
\\
\label{ec3d}
k^2\,\theta_2^{\ 2}
+\alpha\,Z_2
&=&
-\alpha\,\frac{f}{4}\left(2\,\nu''+\nu'^2+2\frac{\nu'}{r}\right)
\nonumber
\\
&&
-\alpha\,\frac{f'}{4}\left(\nu'+\frac{2}{r}\right)
\ ,
\end{eqnarray}
where 
\begin{eqnarray}
\label{Z1}
Z_1
&=&
\frac{e^{-\mu}\,g'}{r}
\\
\label{Z2}
4\,Z_2&=&e^{-\mu}\left(2g''+g'^2+\frac{2\,g'}{r}+2\xi'\,g'-\mu'g'\right)
\ .
\end{eqnarray}
The above equations clearly show that the tensor $\theta_{\mu\nu}$ must vanish
when the deformations vanish ($\alpha=0$).
Moreover, on assuming $g=0$, Eqs.~\eqref{ec1d}-\eqref{ec3d} reduce to the simpler
``quasi-Einstein" system of the MGD of Ref.~\cite{Ovalle:2017fgl},
in which $f$ is only determined by $\theta_{\mu\nu}$ and the undeformed metric~\eqref{pfmetric}.
\par
It is also important to discuss the conservation equation~\eqref{dT0} which now reads
\begin{eqnarray}
\label{con111}
&&
\left({T}_1^{\ 1}\right)'
-
\frac{\xi'}{2}\left({T}_0^{\ 0}-{T}_1^{\ 1}\right)
-
\frac{2}{r}\left({T}_2^{\ 2}-{T}_1^{\ 1}\right)
\nonumber
\\
&&
-\frac{\alpha\,g'}{2}\left({T}_0^{\ 0}-{T}_1^{\ 1}\right)
\nonumber
\\
&&
+\left({\theta}_1^{\ 1}\right)'
-
\frac{\nu'}{2}\left({\theta}_0^{\ 0}-{\theta}_1^{\ 1}\right)
-
\frac{2}{r}\left({\theta}_2^{\ 2}-{\theta}_1^{\ 1}\right)
=
0
\ .
\quad
\end{eqnarray}
The first line precisely represents the divergence of $T_{\mu\nu}$ computed with the
covariant derivative $\nabla^{(\xi,\mu)}$ for the metric~(\ref{pfmetric}),
and is a linear combination of the Einstein field equations~\eqref{ec1pf}-\eqref{ec3pf}.
In fact, the Einstein tensor ${G}_{\mu\nu}$ for the metric~(\ref{pfmetric}) must satisfy its respective
Bianchi identity, and the energy momentum tensor $T_{\mu \nu }$ is therefore conserved
by construction in this geometry,
\begin{equation}
\label{pfcon}
\nabla^{(\xi,\mu)}_\sigma\,T^{\sigma}_{\ \nu}=0
\ ,
\end{equation}
Upon using the deformed metric in Eq.~\eqref{metric}, one instead obtains
\begin{equation}
\label{divs}
\nabla_\sigma\,T^{\sigma}_{\ \nu}
=
\nabla^{(\xi,\mu)}_\sigma\,T^{\sigma}_{\ \nu}
-
\alpha\,\frac{g'}{2}\left({T}_0^{\ 0}-{T}_1^{\ 1}\right)\delta^1_\nu
\ ,
\end{equation}
which explains the origin of the term in the second line of Eq.~\eqref{con111}.
Finally, Eq.~\eqref{con111} becomes
\begin{eqnarray}
\label{con22}
0
&=&
\nabla_\sigma\,T^{\sigma}_{\ \nu}
+\nabla_\sigma\theta^{\sigma}_{\ \nu}
\nonumber
\\
&=&
\alpha\,\frac{g'}{2}\left({T}_0^{\ 0}-{T}_1^{\ 1}\right)\delta^1_\nu
+
\nabla_\sigma\theta^{\sigma}_{\ \nu}
\nonumber
\\
&=&
\left({\theta}_1^{\ 1}\right)'
-
\frac{\nu'}{2}\left({\theta}_0^{\ 0}-{\theta}_1^{\ 1}\right)
-
\frac{2}{r}\left({\theta}_2^{\ 2}-{\theta}_1^{\ 1}\right)
\nonumber
\\
&&
-
\alpha\,\frac{g'}{2}\left({T}_0^{\ 0}-{T}_1^{\ 1}\right)
\ ,
\end{eqnarray}
which is also a linear combination of the ``quasi-Einstein'' field equations~\eqref{ec1d}-\eqref{ec3d} 
for the source $\theta_{\mu\nu}$.
We therefore conclude that the two sources $T_{\mu\nu}$ and $\theta_{\mu\nu}$ can be successfully decoupled
by means of the EGD.
This result is particularly remarkable, since it does not require a perturbative expansion in the parameter 
$\alpha$~\cite{Ovalle:2020fuo}.
\section{Hairy black holes}
\label{Shbh}
Conditions to evade the no-hair theorem have been investigated for a long time~\cite{Sotiriou:2011dz,Babichev:2013cya,Cisterna:2014nua,Sotiriou:2013qea,
	Antoniou:2017acq,Antoniou:2017hxj,Grumiller:2019fmp,Volkov:1989fi,Kanti:1995vq,Kanti:1997br,Zloshchastiev:2004ny}. A straightforward possibility is to fill the static vacuum with some source of potentially fundamental origin,
often described as a scalar field~\cite{Martinez:2004nb,Herdeiro:2015waa,Sotiriou:2015pka}.
We recently considered a more general scenario within the MGD approach,
where the Schwarzschild vacuum for $T_{\mu\nu}=0$ is filled with a generic static and
spherically symmetric source of energy-momentum tensor $\theta_{\mu\nu}$,
namely, a ``tensor-vacuum''~\cite{Ovalle:2018umz}.
This leads to hairy black hole solutions with a rich geometry described by the mass $M$ and
a discrete set of charges generating primary hair.
However, the MGD~\eqref{gd2} leaves the temporal component of the metric~\eqref{metric} 
exactly equal to the Schwarzschild one, 
\begin{equation}
e^\nu
=
e^\xi
=
1-\frac{2\,M}{r}
\ ,
\label{schwR}
\end{equation}
which hinders the existence of stable black holes with a well-defined event horizon.
Indeed, the relation~\eqref{schwR} implies that only hairy black hole solutions 
with the event horizon at $r_{\rm H} = 2\,M$ can be free of pathologies.
The advantage of the EGD is that the temporal component is also modified
according to Eq.~\eqref{gd1}, thus yielding a potentially larger number of hairy black
hole solutions with horizons other than $r_{\rm H} = 2\,M$.
\par
We apply the analysis in the previous Section to the particular case of $T_{\mu\nu}=0$.
The seed metric~\eqref{pfmetric} is thus given by the Schwarzschild solution with 
\begin{equation}
e^\xi
=
e^{-\mu}
=
1-\frac{2\,M}{r}
\ ,
\label{schw}
\end{equation}  
which solves Eqs.~\eqref{ec1pf}-\eqref{ec3pf} for $T_{\mu\nu}=0$.
In order to find hairy black holes, we then need to solve the resulting
``quasi-Einstein'' system~\eqref{ec1d}-\eqref{ec3d}, which contain the three components
of $\theta_{\mu\nu}$ and the two deformations $f$ and $g$.
Furthermore, we reduce the number of unknown quantities, so that they specify
a unique solution, by prescribing the two conditions discussed in the Introduction. 
\par
First of all, in order to have black hole solutions with a well-defined horizon structure,
we demand that the deformed metric~\eqref{metric} satisfies~\footnote{We remark that
Eq.~\eqref{constr1} implies the condition~\eqref{cRh} but is not necessary for it to hold,
as one could also consider cases with $e^{\nu}\not= e^{-\lambda}$ for $r\not=r_{\rm H}$.}
\begin{equation}
\label{constr1}
e^{\nu}=e^{-\lambda}
\ .
\end{equation}
This condition ensures that the radius $r=r_{\rm H}$ such that
\begin{equation}
\label{cRh}
e^{\nu(r_{\rm H})}=e^{-\lambda(r_{\rm H})}=0
\end{equation}
will be both a killing horizon ($e^{\nu}=0$) and a causal horizon $(e^{-\lambda}=0)$.
A direct consequence of the condition~\eqref{constr1}, following from the Einstein
equations~\eqref{ec1} and~\eqref{ec2}, is the equation of state
\begin{equation}
\tilde{p}_{r}
=
-\tilde{\rho}
\ .
\label{schwcon}
\end{equation}
Therefore, only negative radial pressure is allowed (for positive density).
The condition~\eqref{constr1} and the Schwarzschild solution~\eqref{schw} then relate the
metric deformations $f$ and $g$ according to
\begin{equation}
\label{fg}
\alpha\,f(r)
=
\left(1-\frac{2\,M}{r}\right)\left(e^{\alpha\,g(r)}-1\right)
\ ,
\end{equation}
so that the line element~\eqref{metric} becomes
\begin{eqnarray}
\label{hairyBH}
ds^{2}
&=&
\left(1-\frac{2\,M}{r}\right)
e^{\alpha\,g(r)}
dt^{2}
-\left(1-\frac{2\,M}{r}\right)^{-1}
e^{-\alpha\,g(r)}
dr^2
\nonumber
\\
&&
-r^{2}\,d\Omega^2
\ .
\end{eqnarray}
\par
We are now left with the deformation $g$ and the three components of $\theta_{\mu\nu}$,
which must satisfy the three ``quasi-Einstein'' Eqs.~\eqref{ec1d}-\eqref{ec3d}.
We can therefore impose some physically motivated restriction on $g$ or a constraint on
$\theta_{\mu\nu}$, like a reasonable equation of state.
For instance, in the region $r\geq\,2\,M$, we can consider the tensor-vacuum satisfies 
\begin{equation}
\label{generic}
\theta_0^{\,0}
=
a\,\theta_1^{\,1}+b\,\theta_2^{\,2}
\ ,
\end{equation}
with $a$ and $b$ constants. 
In this case, Eqs.~(\ref{ec1d})-(\ref{ec3d}) yield the differential equation 
\begin{eqnarray}
\label{master}
&&b\,r\,(r-2\,M)\,h''+2\,\left[(a+b-1)\,r-2\,(a-1)\,M\right]
h'
\nonumber
\\
&&
+2\,(a-1)\,h=2\,(a-1)
\ ,
\end{eqnarray}
for 
\begin{equation}
h(r)
=
e^{\alpha\,g(r)}
\ .
\label{h}
\end{equation}
The solution can be written as
\begin{equation}
\label{master2}
e^{\alpha\,g(r)}
=
1+\frac{1}{r-2\,M}
\left[\ell_0+r\left(\frac{\ell}{r}\right)^{2(1-a)/b}
\right]
\ ,
\end{equation}
where $\ell_0$ and $\ell$ are constants (proportional to $\alpha$) with dimensions of a length.
By using this expression in the line element~\eqref{hairyBH},
we obtain the metric functions
\begin{equation}
\label{bh}
e^{\nu}
=
e^{-\lambda}
=
1-\frac{2\,{\cal M}}{r}
+
\frac{{\ell}^n}{r^n}
\ ,
\end{equation}
where we defined the new mass as ${\cal M}=M+\ell_0/2$ and
\begin{equation}
\label{n}
n
=
\frac{2}{b}
\left(a-1\right)
\ ,
\end{equation} 
with $n>1$ for a correct asymptotic behavior.~\footnote{Note that a trivial deformation
which leaves the Schwarzschild metric~\eqref{schw} unaffected is also recovered by
setting $a=1$ ($n=0$) in Eq.~\eqref{master}.
We will have more to say about this in the next subsections.}
The possible horizons $r=r_{\rm H}$ are given by the solutions of
\begin{equation}
\label{horizon}
r_{\rm H}^n-2\,{\cal M}\,r_{\rm H}^{n-1}+\ell = 0
\ ,
\end{equation}
and the space-time represents a Kiselev black hole~\cite{Kiselev:2002dx},
which was analyzed in great detail by Visser in Ref.~\cite{Visser:2019brz}. 
This line element is produced by the effective density 
\begin{equation}
\tilde{\rho}
=
\theta_0^{\ 0}
=
\alpha\,\frac{(n-1)\,\ell^n}{k^2\,r^{n+2}}
\ ,
\label{efecdenx}
\end{equation}
the effective radial pressure
\begin{equation}
\tilde{p}_{r}
=
-\theta_1^{\ 1}
=
-\tilde{\rho}
\ ,
\label{efecprax}
\end{equation}
and the effective tangential pressure
\begin{equation}
\tilde{p}_{t}
=
-\theta_2^{\ 2}
=
\frac{n}{2}\,\tilde{\rho}
\ .
\label{efecptanx}
\end{equation}
The anisotropy~\eqref{anisotropy} is thus given by
\begin{equation}
\Pi
=
\left(\frac{n}{2}+1\right)\,\tilde{\rho}
\ .
\end{equation}
We see that the Schwarzschild-de~Siter solution ($n=-2$) is the only one which allows
for an isotropic tensor-vacuum.
On the other hand, the DEC, namley $\tilde{\rho}\geq |\tilde{p}_r|$ and
$\tilde{\rho}\geq|\tilde{p}_t|$, yields $n\le2$.
Combining this with asymptotic flatness ($n>1$), we obtain the range
\begin{equation}
\label{n}
1
\le
n
\le
2
\ .
\end{equation}
The extreme cases $n=1$ and $n=2$ are, respectively, the Schwarzschild solution and the
conformal solution with traceless $\theta_{\mu\nu}$, like the Maxwell case.
Since the Kiselev black hole has already been studied extensively, we will abandon Eq.~\eqref{generic}
and continue analysing the deformed metric~\eqref{hairyBH} based on energy conditions.
\par
Let us recall that the energy conditions are a set of requirements which are usually imposed
on the energy-momentum tensor to avoid exotic matter sources, hence we can see them
as sensible guidelines to avoid classically unphysical configurations~\cite{Visser:1995cc,Curiel:2014zba}.
In particular, we will impose energy conditions on the source $\theta_{\mu\nu}$ in the
region of space-time accessible to an outer observer (while possibly relax them
inside the event horizon).
\subsection{Strong energy condition}
\label{SSsec}
Let us start with the SEC, namely
\begin{eqnarray}
\nonumber
&&
\tilde{\rho}+\tilde{p}_r+2\,\tilde{p}_t
\geq
0
\\
\label{strong01}
&&
\tilde{\rho}+\tilde{p}_r
\geq
0
\\
&&
\tilde{\rho}+\tilde{p}_t
\geq
0
\ ,
\nonumber
\end{eqnarray}
which, as a consequence of Eq.~\eqref{schwcon}, reduce to
\begin{eqnarray}
\label{strong2}
\theta_2^{\ 2}
&\leq&
0
\\
\theta_0^{\ 0}
&\geq&
\theta_2^{\ 2}
\ ,
\label{strong3}
\end{eqnarray}
where we used the definitions~\eqref{efecden} and~\eqref{efecpretan}.
\par
Using Eqs.~\eqref{ec1d} and~\eqref{ec3d} we find that the conditions~\eqref{strong2}
and~\eqref{strong3} respectively lead to the second-order linear differential inequalities
\begin{eqnarray}
\label{strong5}
&&G_1(r)\equiv
{(r-2\,M) \,h''+2\,h'}
\geq
0\ ,\\
\label{strong51}
&&G_2(r)
\equiv
r\,(r-2\,M)\,h''
+4\,M\,h'
-2\,h+2
\geq
0\ ,
\end{eqnarray} 
where $h$ was defined in Eq.~\eqref{h}.
It is now useful to recall that a Gr\"{o}nwall's inequality of the form
\begin{equation}
\label{gro1}
U'(r)\leq\,\beta(r)\,U(r)
\ ,
\end{equation}
in the interval $r_0\leq\,r\,\leq\,\infty$, admits the solution 
\begin{equation}
\label{gro2}
U(r)
\leq
U(r_0)\,
e^{\int_{r_0}^{r}\beta(s)\,ds}
\equiv
U_0(r)
\ ,
\end{equation}
where the bounding function $U_0$ is obtained by saturating the differential inequality~\eqref{gro1}.
For the inequality~\eqref{strong5}, we can define 
\begin{equation}
U(r)=-h'(r)
\end{equation}
and $\beta(r)=-2\,(r-2\,M)^{-1}$, so that Eq.~\eqref{gro2} yields
\begin{equation}
h'(r)\geq\,h'(r_0)\left(\frac{r_0-2\,M}{r-2\,M}\right)^2
\ ,
\end{equation}
and finally
\begin{equation}
h(r)
\geq
h(r_0)
+
(r_0-2\,M)
\left(1-\frac{r_0-2\,M}{r-2\,M}\right)
h'(r_0)
\ .
\end{equation}
We therefore find that the bounding function solving $G_1(r)=0$ behaves as
\begin{equation}
\label{beh}
h_1(r)
\sim
c_1-\frac{\ell_1}{r-2\,M}
\ ,
\end{equation}
where $c_1$ is a dimensionless constants and $\ell_1$ a constant with dimensions of a length.
However, any deformation of the form in Eq.~\eqref{beh} plugged into the metric~\eqref{hairyBH}
uniquely leads to
\begin{equation}
e^\nu
=
e^{-\lambda}
=
c_1
\left(
1
-\frac{2\,c_1\,M+\ell_1}{c_1\,r}
\right)
\ ,
\end{equation}
which becomes the Schwarzschild solution~\eqref{schw} by imposing asymptotic flatness
(that is, setting $c_1=1$) and rescaling the mass $2\,M+\ell_1\to 2\,M$.
Indeed, we notice that $G_1(r)=0$ equals the differential equation~\eqref{master} for $a=1$,
whose solution yields the Schwarzschild metric.
On the other hand, following the same procedure for the inequality~\eqref{strong51},
we find that the bounding function satisfying $G_2(r)=0$ behaves as
\begin{equation}
\label{beh2}
h_2(r)
\sim
1-\frac{1}{r-2\,M}
\left(\ell_2
-\frac{\Lambda}{3}\,{r^3}
\right)
\ ,
\end{equation}
where $\ell_2$ is again a length and $\Lambda$ a constant with dimensions of the inverse
of a squared length.
Likewise, the deformation~\eqref{beh2} plugged into the metric~\eqref{hairyBH} leads to
\begin{equation}
e^\nu
=
e^{-\lambda}
=
1
-\frac{2\,M+\ell_2}{r}+\frac{\Lambda}{3}\,r^2
\ ,
\end{equation}
which is the Schwarzschild-de~Sitter metric with cosmological constant $\Lambda$ and mass
$2\,M+\ell_2\to 2\,M$.
Again, notice that $G_2(r)=0$ is the differential equation~\eqref{master} for $a=0$ and $b=1$
and the bounding solution $h_2$ for the extremal case $G_2(r)=0$ (with $\alpha\neq 0$) leads
to the line element~\eqref{bh} with $n=-2$.
Since both inequalities~\eqref{strong5} and~\eqref{strong51} must hold, the unique bounding
deformation $h_0(r)$ which solves $G_1(r)=G_2(r)=0$ is obtained when Eq.~\eqref{beh}
equals Eq.~\eqref{beh2}, that is for $\Lambda=0$.
We thus conclude that the bounding deformations which saturate the SEC are of the form
\begin{equation}
h_0(r)
=
1
-\frac{\ell_0}{r-2\,M}
\label{h0sec}
\end{equation}
and leave the Schwarzschild geometry unaffected.
This is not at all surprising since $G_1(r)=G_2(r)=0$ is tantamount to
$\tilde\rho=\tilde p_r=\tilde p_t=0$ and the Schwarzschild geometry cannot possibly be
deformed in this case.
\par
Before we proceed to consider deformations which do not saturate the inequalities~\eqref{strong5}
and~\eqref{strong51}, we notice that the transformation
\begin{equation}
\label{gauge}
h(r)
\to
\bar h(r)
\equiv
h(r)-\frac{\ell_0}{r-2\,M}
\end{equation}
leaves $G_1(r)$ and $G_2(r)$ invariant.
Under this transformation, the metric functions change as
\begin{equation}
\label{hgauge}
e^\nu
=
e^{-\lambda}
=
\left(1-\frac{2\,M}{r}\right)
h(r)
\to
\left(1-\frac{2\,M}{r}\right)
h(r)
-
\frac{\ell_0}{r}
\ ,
\end{equation}
where it is reasonable to assume that $h(r)$ admits an expansion in powers of $1/r$ for
a regular exterior.
In particular, the effect of the metric transformation~\eqref{hgauge} will amount to
the usual shift of the mass $M\to{\cal M}=M+\ell_0/2$ at order $1/r$.
This redefinition of the asymptotic mass, in turn, will introduce a new dependence on the length
$\ell_0$ in $h(r)$ whenever the latter contains the unshifted seed mass $M={\cal M}-\ell_0/2$,
thus generating a new solution with parameters ${\cal M}$ and $\ell_0$.
Of course, for $h(r)\sim 1$, Eq.~\eqref{gauge} acts like a ``gauge'' symmetry
corresponding to the trivial deformations~\eqref{h0sec} of the seed Schwarzschild
geometry~\eqref{schw}.
\par
We have just seen that the parameter $\ell_0$ in Eq.~\eqref{gauge} appears as a
new ``gauge'' charge.
The way this all works can be made more explicit by considering concrete examples.
Since we are interested in solutions with a proper horizon at $r=r_{\rm H}\sim 2\,M$, which also
behave approximately like the Schwarzschild metric for $r\gg 2\,M$ (so as to meet all experimental
bounds in the weak field regime), we could consider any positive function $G_1(r)$ satisfying the
boundary conditions
\begin{equation}
\label{condition1}
G_1(r)=0
\qquad
{\rm for}
\quad
\left\{
\begin{array}{ll}
r\sim
2\,M
\\
\\
r
\gg
M
\ ,
\end{array}
\right.
\end{equation}
A simple example of such a function containing just the parameters $\alpha$ and $M$ is given by
\begin{equation}
\label{G}
G_{\alpha,M}(r)
=
\frac{\alpha}{M^2}\,(r-2\,M)\,e^{-{r}/{M}}
\ .
\end{equation}
Upon solving Eq.~\eqref{strong5} for the corresponding deformation $h$, we obtain 
\begin{equation}
\label{strongg}
h(r)
=
c_1
-
\alpha\,\frac{\ell-r\,e^{-r/M}}{r-2\,M}
\ ,
\end{equation}
where $\alpha\,\ell=\ell_0$ and we can set $c_1=1$ to recover the proper limit for $\alpha=0$
(in which $G_{\alpha,M}\to 0$). 
The deformation in Eq.~\eqref{strongg} must also satisfy the inequality~\eqref{strong51},
which becomes
\begin{equation}
\label{Gg}
\frac{\alpha}{M^2}
\left(r^2-2\,M^2\right)
e^{-{r}/{M}}
\geq
0
\ ,
\end{equation}
and it is satisfied for all $r\geq\,\sqrt{2}\,M$.
\par
Finally, using~\eqref{strongg} in the line element~\eqref{hairyBH} yields the metric functions
\begin{equation}
\label{strongBH}
e^{\nu}
=
e^{-\lambda}
=
1-\frac{2\,{\cal M}}{r}+\alpha\,e^{-r/({\cal M}-\alpha\,\ell/2)}
\ ,
\end{equation}
where we modded out the term proportional to $\ell$, corresponding to the gauge
transformation~\eqref{gauge}, by introducing the mass ${\cal M}=M+\alpha\,\ell/2$.
The new solution~\eqref{strongBH} thus asymptotically approaches the Schwarzschild
geometry with a total mass ${\cal M}$.
The source $\theta_{\mu\nu}$ decays exponentially away from the center of the system, 
as can be seen from the effective density
\begin{equation}
\tilde{\rho}
=
\theta_0^{\ 0}
=
-\tilde{p}_r
=
\frac{\alpha\,e^{-r/M}}{k^2\,M\,r^2}\left(r-M\right)
\ ,
\label{denstrong}
\end{equation}
and the effective tangential pressure
\begin{equation}
\tilde{p}_{t}
=
-\theta_2^{\ 2}
=
\frac{\alpha\,e^{-r/M}}{2\,k^2\,M^2\,r}\left(r-2\,M\right)
\ .
\label{prestrong}
\end{equation}
We can immediately see some important features of the metric~\eqref{strongBH}.
The first one is that the physical singularity at $r=0$ remains and is further reflected
in the singular behaviour of the effective quantities in Eqs.~\eqref{denstrong}
and~\eqref{prestrong}.
The second feature is that the SEC~\eqref{strong2}
is only satisfied as long as $r\geq 2\,M$, as we can see from~\eqref{prestrong}.
Also notice that in the limit $M\to 0$ the source $\theta_{\mu}^{\ \nu}\to 0$ for $r>0$,
and ${\cal M}\to \alpha\,\ell$.
In other words, the source $\theta_{\mu}^{\ \nu}$ approaches a Dirac-delta function
for vanishing seed mass $M$. 
\begin{figure}[t]
\center
\includegraphics[width=8cm]{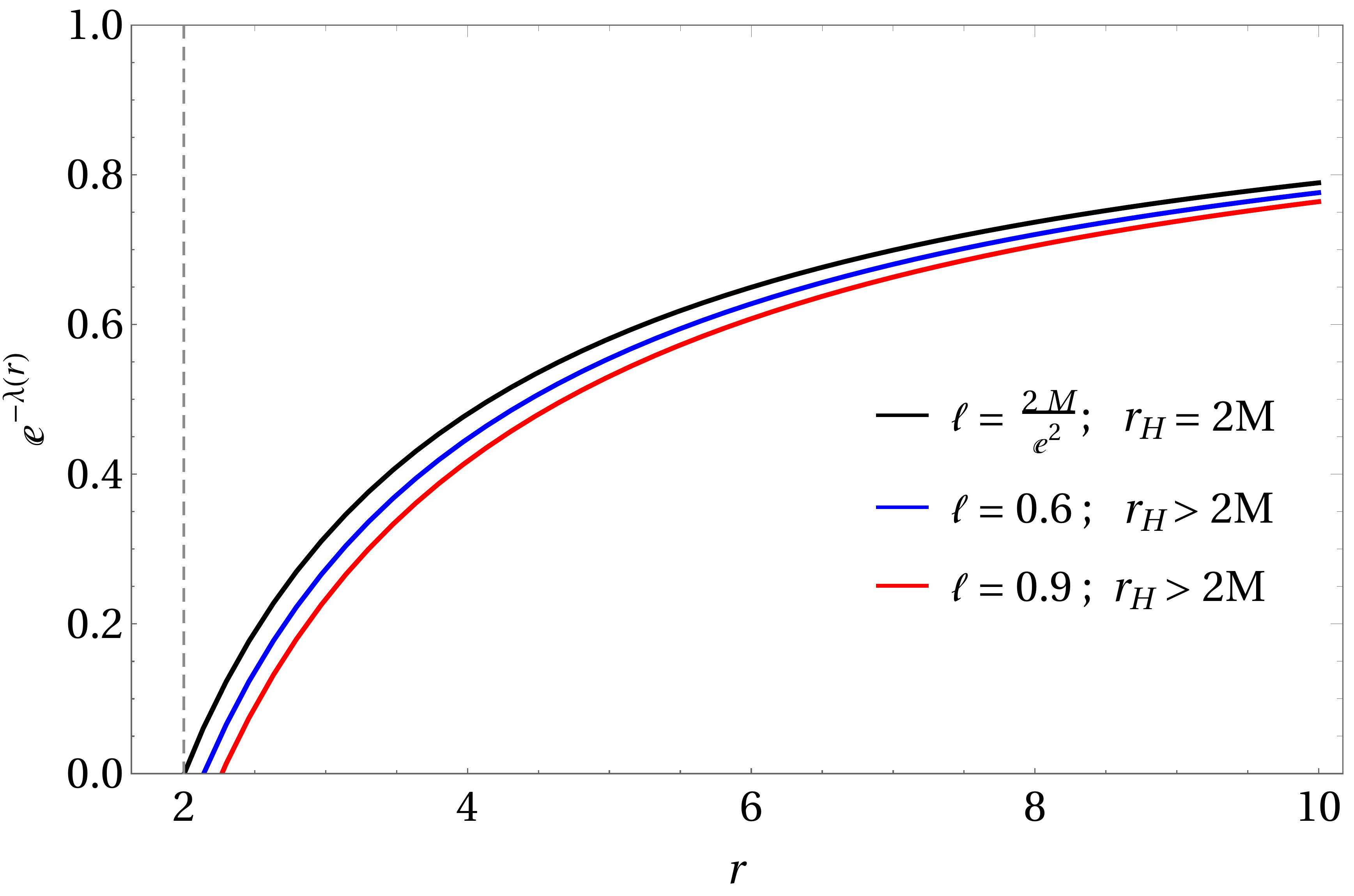}
\\
\caption{Metric function for different values of $\ell$ with $\alpha=0.4$
and $M=1$. 
The shift of the horizon $r_{\rm H}$ is controlled by the parameter $\ell$. The Schwarzschild horizon corresponds to the saturated case of the inequality~\eqref{lineq}.}
\label{fig1}      
\end{figure}
\begin{figure}[t]
\center
\includegraphics[width=8cm]{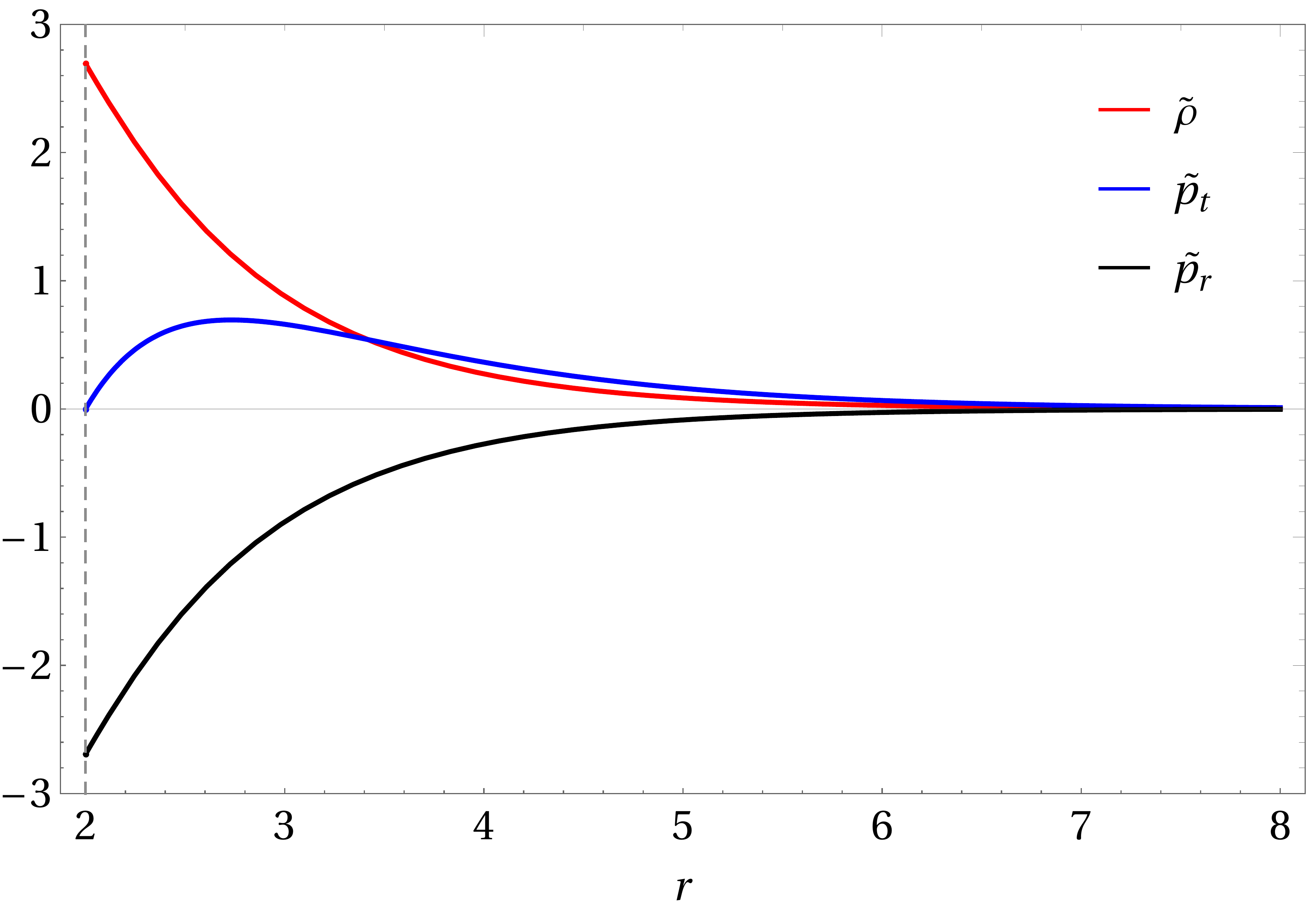}
\\
\caption{Effective source terms $\times\,10^{4}$ for $\alpha=0.2$.
The horizon is located at $r_{\rm H}\gtrsim\,2\,M$, with $M=1$. }
\label{fig2}      
\end{figure}
\par
The equation determining the horizon $r=r_{\rm H}$ of the metric~\eqref{strongBH}
is given by
\begin{equation}
\label{stronghorizon}
\alpha\,\ell
=
r_{\rm H}-2\,M+\alpha\,r_{\rm H}\,e^{-r_{\rm H}/M}
\ ,
\end{equation}
which allows to write the metric~\eqref{strongBH} in terms of its horizon in the form
\begin{eqnarray}
\label{strongBH3}
e^{\nu}
=
e^{-\lambda}
=
1-\frac{\,r_{\rm H}}{r}+\alpha\,\left(e^{-r/M}-\frac{\,r_{\rm H}}{r}\,e^{-r_{\rm H}/M}\right)
\ .
\quad
\end{eqnarray}
It is of course impossible to find analytical solutions to Eq.~\eqref{stronghorizon},
except for particular values of the parameters.
For example, according to our prescription for the SEC, we need
$r_{\rm H}\geq 2\,M$, or
\begin{equation}
\label{lineq}
\ell
\geq
2\,M/e^{2}
\ .
\end{equation}
The extremal case $\ell=2\,M/e^{2}$ leads to the solution
\begin{eqnarray}
\label{strongh2M}
e^{\nu}
=
e^{-\lambda}
=
1-\frac{2\,M}{r}+\alpha\,\left(e^{-r/M}-\frac{2\,M}{e^2\,r}\right)
\ .
\end{eqnarray}
which has the horizon at $r_{\rm H} = 2\,{M}$.
The SEC~\eqref{strong2} and~\eqref{strong3} are both satisfied
in the outer region, but the black hole should have the same thermodynamic properties
of the Schwarzschild geometry.
\par
The metric function~\eqref{strongBH} is plotted in Fig.~\ref{fig1} for a given value of $\alpha$,
which shows that the horizon is shifted to larger radii when $\ell$ increases from the minimum
allowed value~\eqref{lineq} corresponding to $r_{\rm H} = 2\,{M}$.
We find the same behavior when $\alpha$ increases.
In this respect, we notice that the effective density ${\tilde \rho}$ and pressures
${\tilde p}_r$ and ${\tilde p}_t$ in Eqs.~\eqref{denstrong} and~\eqref{prestrong} do not
depend on the parameter $\ell$, unlike the horizon.
This allows us to choose suitable values for the parameters $\alpha$ and $\ell$ such that
the SEC is satisfied for $r\gtrsim r_{\rm H}$, as is shown in Fig.~\ref{fig2},
where both the density and tangential pressure are positive.
We conclude that the metric~\eqref{strongBH} represents a hairy black hole
(having in general non-trivial thermodynamic properties) endowed with the parameters
$\{M,\alpha,\ell\}$, where $\ell_0=\alpha\,\ell$ represents a charge associated with
primary hair.
%
%
%
\subsection{Dominant energy condition}
\label{SSdec}
We shall next consider the DEC, which requires~\cite{Visser:1995cc}
\begin{eqnarray}
\tilde{\rho}
&\geq&
|\tilde{p}_r|
\label{dom1}
\\
\tilde{\rho}
&\geq&
|\tilde{p}_t|
\ .
\label{dom2} 
\end{eqnarray}
In particular, we will see that these conditions allow for deforming the Schwarzschild
solution into the Reissner-Nordstr\"{o}m-de~Sitter geometry with an effective charge $Q$
and an effective cosmological constant $\Lambda$.
\par
We first point out that the inequality~\eqref{dom1} is saturated as a consequence
of Eq.~\eqref{schwcon} for a positive effective density, for which Eq.~\eqref{dom2}
reduces to
\begin{equation}
\label{dom3}
-\tilde{\rho}
\le
\tilde{p}_t
\le
\tilde{\rho}
\ .
\end{equation}
We can again write the condition~\eqref{dom3} in terms of the
definitions~\eqref{efecden} and~\eqref{efecpretan} as
\begin{eqnarray}
\label{dom4}
&&
\theta_0^{\ 0}+\,\theta_2^{\ 2}
\geq
0
\\
\label{dom41}
&&
\theta_0^{\ 0}-\,\theta_2^{\ 2}
\geq
0
\ ,
\end{eqnarray}
which yield respectively the differential inequalities
\begin{eqnarray}
\label{dom6}
&&H_1(r)
\equiv
-r(r-2\,M)h''
-4(r-M)h'
-2h+2
\geq
0
\qquad
\\
\label{dom61}
&&H_2(r)
\equiv
r\,(r-2\,M)\,h''
+4\,M\,h'
-2\,h+2
\geq
0
\ ,
\end{eqnarray} 
where we used Eqs.~\eqref{ec1d} and~\eqref{ec3d}, and $h$ is the same defined in Eq.~\eqref{h}.
\par
We can then notice that $H_1(r)=0$ is the differential equation~\eqref{master} for $a=0$ and $b=-1$.
Hence, the bounding solution $h_1$ for the extremal case $H_1(r)=0$ (with $\alpha\neq\,0$) leads
to the line element~\eqref{bh} with $n=2$, namely, the Reissner-Nordstr\"om solution
(with an effective charge $Q\sim\ell$).
On the other hand, $H_2(r)=0$ is the differential equation~\eqref{master} for $a=0$ and $b=1$.
Hence, the bounding solution $h_2$ for the extremal case $H_2(r)=0$ leads
to the line element in~\eqref{bh} with $n=-2$, namely, the Schwarzschild-de~Sitter solution with
cosmological constant $\Lambda\sim\ell^{-2}$.
Since $H_1(r)=H_2(r)=0$ corresponds to vanishing $\theta_{\mu\nu}$ like for the SEC,
the unique bounding deformation $h_0(r)=h_1(r)=h_2(r)$ is obtained for $Q=\Lambda=0$,
so that the only possible deformation remains again the trivial one in Eq.~\eqref{h0sec},
which yields the seed Schwarzschild solution. 
Indeed, the functions $H_1(r)$ and $H_2(r)$ in Eqs.~\eqref{dom6} and~\eqref{dom61}
are also invariant under the transformation~\eqref{gauge}.
\par 
Like in Section~\ref{SSsec}, we proceed to investigate deformations which do not saturate
the inequalities~\eqref{dom6} and~\eqref{dom61} everywhere by considering positive
functions $H_1(r)$ or $H_2(r)$ which saturate that inequalities only near the boundaries
of the outer region, that is
\begin{equation}
H_1(r)=0
\qquad
{\rm for}
\quad
\left\{
\begin{array}{ll}
r\sim
2\,M
\\
\\
r
\gg
M
\ .
\end{array}
\right.
\end{equation}
In fact, we can still employ the function in Eq.~\eqref{G} and set
\begin{equation}
\label{H}
H_1=M\,G_{\alpha,M}
\ .
\end{equation}
Upon solving~\eqref{dom6} for the corresponding $h$, we obtain 
\begin{equation}
\label{dominantg}
h(r)
=
1
-
\frac{1}{r-2\,M}
\left(
\alpha\,\ell
+\alpha\,M\,e^{-r/M}
-\frac{Q^2}{r}
\right)
\ ,
\end{equation}
where $\alpha\,\ell=\ell_0$ and $Q$ is also a constant with dimension of a length and proportional
to $\alpha$.
A second constant of integration was adjusted to meet the proper Schwarzschild limit for
$\alpha\to 0$ (in which we remark that $Q\sim\alpha$ vanishes as well).
The deformation in Eq.~\eqref{dominantg} also has to satisfy the inequality~\eqref{dom61},
which reads
\begin{equation}
\frac{4\,Q^2}{r^2}
\ge
\frac{\alpha}{M}\,(r+2\,M)\,e^{-r/M}
\ .
\end{equation}

Using~\eqref{dominantg} in the line element~\eqref{hairyBH}, we obtain the metric functions
\begin{equation}
\label{dominantBH}
e^{\nu}
=
e^{-\lambda}
=1-\frac{2\,M+\alpha\,\ell}{r}
+\frac{Q^2}{r^2}
-\frac{\alpha\,M\,e^{-r/M}}{r}
\ ,
\end{equation}
which is a sort of ``charged'' version of the solution~\eqref{strongBH} again
with asymptotic mass ${\cal M}=M+\alpha\,\ell/2$.
The effective density is now given by
\begin{equation}
\tilde{\rho}
=
\theta_0^{\ 0}
=
-\tilde{p}_r
=
\frac{Q^2}{k^2\,r^4}
-\frac{\alpha\,e^{-r/M}}{k^2\,r^2}
\label{dendom}
\end{equation}
and an effective tangential pressure reads
\begin{equation}
\tilde{p}_{t}
=
-\theta_2^{\ 2}
=
\frac{Q^2}{k^2\,r^4}
-\frac{\alpha\,e^{-r/M}}{2\,k^2\,M\,r}
\ .
\label{predom}
\end{equation}
We can see that
\begin{equation}
\label{domdom2}
\tilde{\rho}-\tilde{p}_{t}
=
\frac{\alpha\,e^{-r/M}}{2\,k^2\,M\,r^2}(r-2\,M)
\end{equation}
and the DEC is satisfied for $r\geq\,2\,M$, as we originally required.
We can also see that the physical singularity at $r=0$ remains.
\par
The horizon radii $r_{\rm H}$ are given by solutions of
\begin{equation}
\label{horizondom}
\alpha\,\ell
=
r_{\rm H}
-2\,M
+\frac{Q^2}{r_{\rm H}}
-\alpha\,M\,e^{-r_{\rm H}/M}
\ ,
\end{equation}
which allows us to write the metric functions~\eqref{dominantBH} as
\begin{eqnarray}
\label{dominantBHH}
e^{\nu}
=
e^{-\lambda}
&=&
1-\frac{r_{\rm H}}{r}
\left(1+\frac{Q^2}{r_{\rm H}^2}-\frac{\alpha\,M}{r_{\rm H}}\,e^{-r_{\rm H}/M}\right)
\nonumber
\\
&&
+\frac{Q^2}{r^2}-\frac{\alpha\,M}{r}\,e^{-r/M}
\ .
\end{eqnarray}
As with the SEC, it is always possible to choose suitable values
for the parameters $\alpha$, $\ell$ and $Q$ such that analytical solutions for $r_{\rm H}$
can be found.
However, since the DEC requires $r_{\rm H}\geq 2\,M$,
the choice of these values cannot be arbitrary. 
We can see this by evaluating the density~\eqref{dendom} at the horizon,
in addition to using the expression~\eqref{horizondom}, which yields 
\begin{equation}
\label{condi}
Q^2
\geq
4\,\alpha\,(M/e)^2
\quad
{\rm and}
\quad
\ell
\geq
M/e^2
\ .
\end{equation}
\par
We remark that $Q$ does not need to be an electric charge.
It could be, for instance, a tidal charge of extra-dimensional origin or any other source.
However, when $Q$ represents an electric charge, we can say that the electro-vacuum
of the Reissner-Nordstr\"om geometry also contains a tensor-vacuum whose components
are those explicitly proportional to $\alpha$ in Eqs.~\eqref{dendom} and~\eqref{predom}.
Let us recall that the Reissner-Nordstr\"{o}m metric has two horizons:
the event horizon 
\begin{equation}
\label{domhor}
r_{\rm H}
=
M+\sqrt{M^2-Q^2}
<
2\,M
\ ,
\end{equation}
and an internal Cauchy horizon given by
\begin{equation}
\label{cauchy}
r_{\rm CH}
\equiv
M-\sqrt{M^2-Q^2}
<
r_{\rm H}
\ .
\end{equation}
For our solution~\eqref{dominantBH}, we can identify at least three cases for which
the event horizon $r_{\rm H}$ has simple analytical expressions, and the DEC~\eqref{dom1}
and~\eqref{dom2} are satisfied.
As in the Reissner-Nordstr\"{o}m metric, each one of these cases has an internal
Cauchy horizon $r_{\rm CH}<r_{\rm H}$. 
%
%
\subsubsection*{Case~1}
Let us start by considering the case saturating the inequalities~\eqref{condi},
for which the metric components~\eqref{dominantBHH} become
\begin{equation}
\label{dominantBH2}
e^{\nu}
=
e^{-\lambda}
=
1-\frac{2\,M}{r}\left(1+\frac{\alpha}{2\,e^2}\right)
+\frac{4\,\alpha\,M^2}{e^2\,r^2}
-\frac{\alpha\,M}{r}e^{-r/M}
\ .
\end{equation}
The event horizon is again precisely at $r_{\rm H}=2\,M$, which parallels the case of
Eq.~\eqref{strongh2M}, and we can also write 
\begin{equation}
r_{\rm H}
=
\frac{e}{\sqrt{\alpha}}\,Q
\ .
\end{equation}
Notice that by defining
\begin{equation}
{\cal M}=M\left(1+\frac{\alpha}{2\,e^2}\right)
\ ,
\end{equation}
the metric functions~\eqref{dominantBH2} can be written in a more suggestive form as 
\begin{equation}
\label{dominantBH2b}
e^{\nu}
=
e^{-\lambda}
=
1-\frac{2\,{\cal M}}{r}
+\frac{Q^2}{r^2}
-\frac{\sqrt{\alpha}\,Q}{2\,r}\,e^{1-2\sqrt{\alpha}\,r/e\,Q}
\ ,
\end{equation}
which can be interpreted as a nonlinear electrodynamics coupled with gravity. 
\subsubsection*{Case 2}
\begin{figure}[t]
\center
\includegraphics[width=8cm]{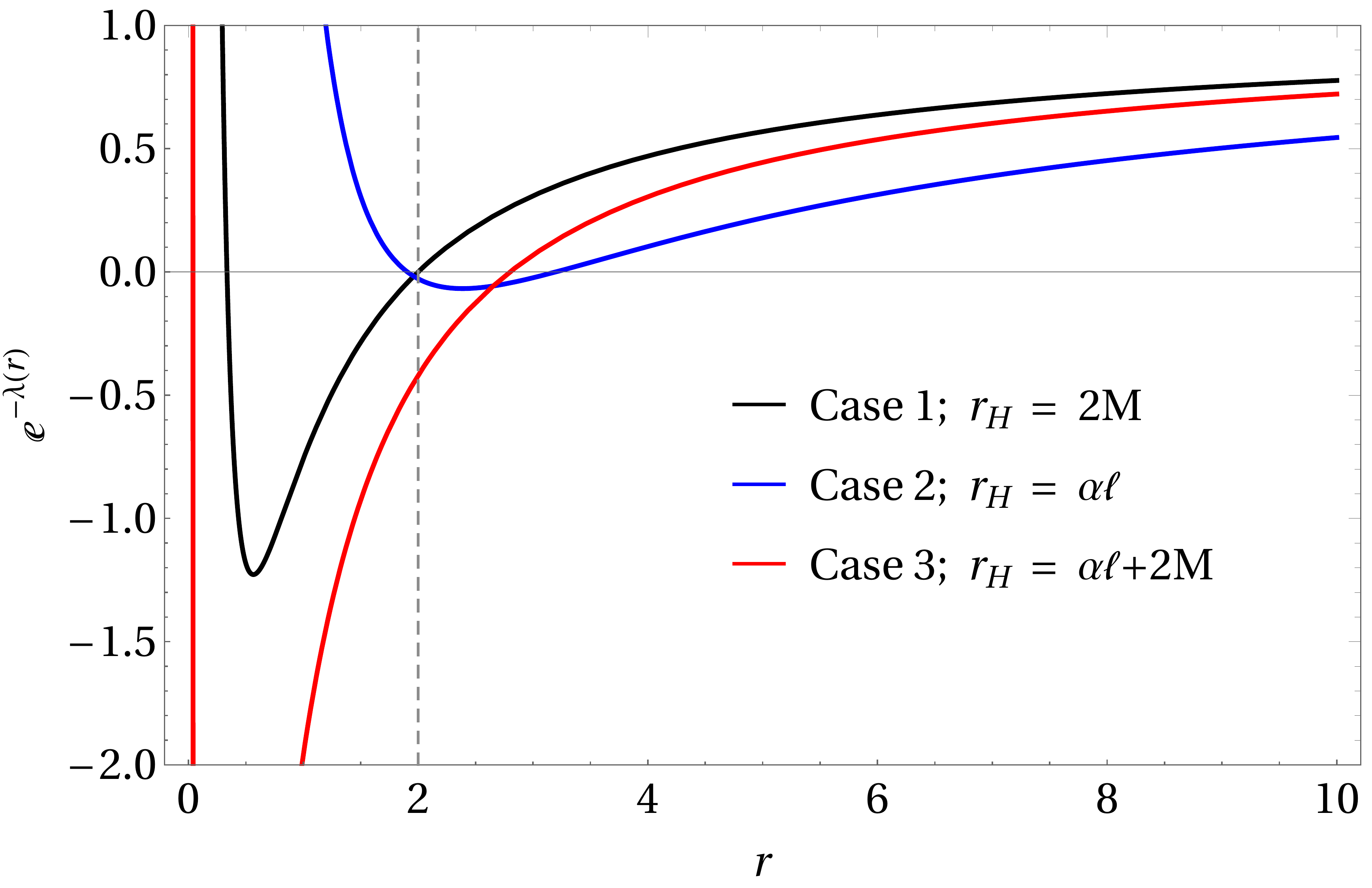}
\\
\caption{Metric function for the three different analytic cases with $\alpha=0.8$ and $M=1$
for $\ell=0.4$, $4$, $1$, respectively. }
\label{fig3}      
\end{figure}
\begin{figure}[t]
\center
\includegraphics[width=8cm]{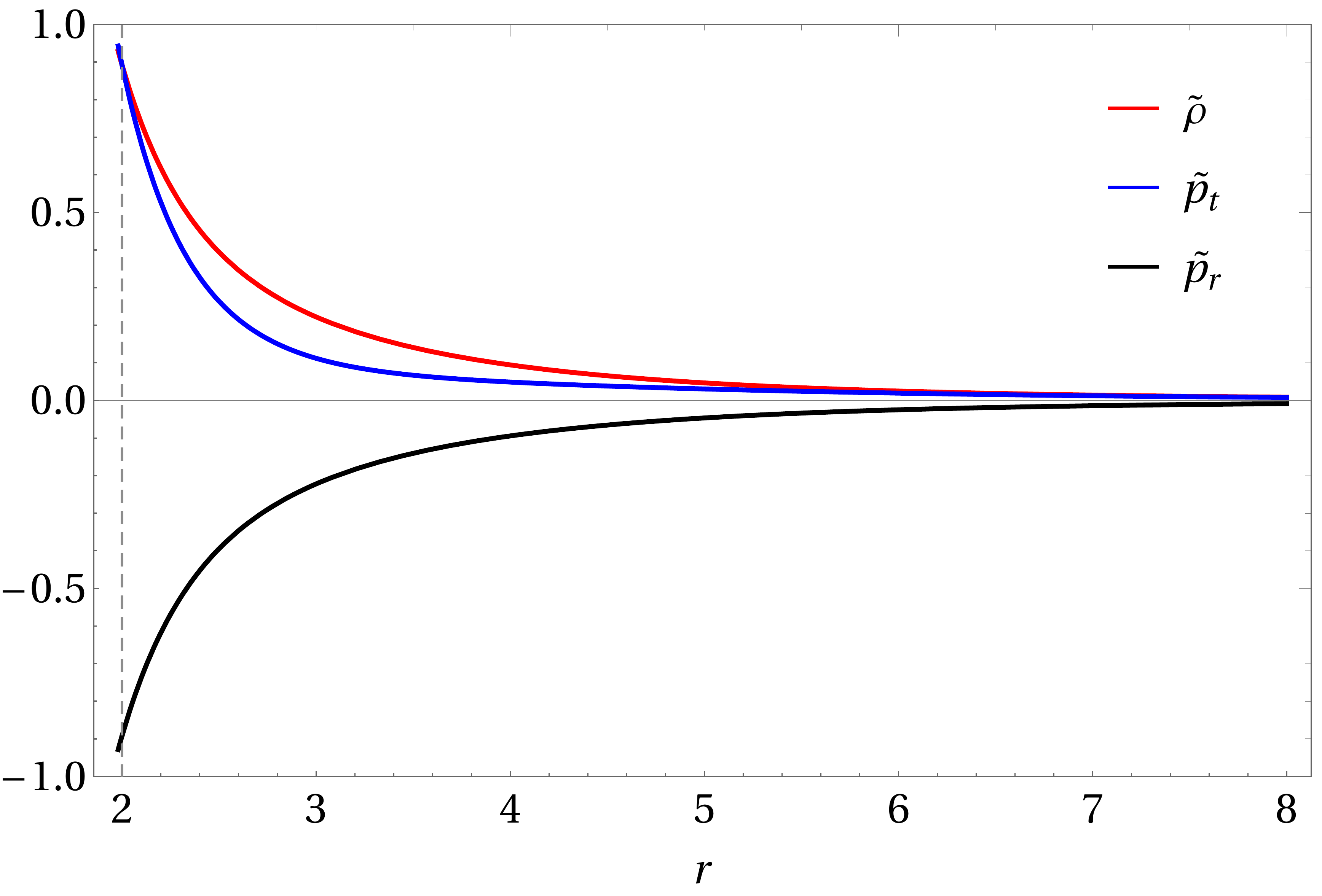}
\\
\caption{Effective source terms $\times 10^{4}$ for $\alpha=0.1$, $\ell=0.4$, $Q=0.3$,
with $M=1$.
}
\label{fig4}      
\end{figure}
The second case we consider is given by
\begin{equation}
\label{qcase2}
Q^2
=
\alpha\,\ell\,M
\left(2+\alpha\,e^{-{\alpha\,\ell}/{M}}\right)
\ ,
\end{equation}
with $\ell\neq 0$, leading to
\begin{eqnarray}
\label{dominantBH3}
e^{\nu}
=
e^{-\lambda}
&=&
1-\frac{2\,M+\alpha\,\ell}{r}
+\frac{2\,\alpha\,\ell\,M}{r^2}
\nonumber
\\
&&
-\frac{\alpha\,M}{r^2}\,e^{-r/M}\left(r-\alpha\,\ell\,e^{\frac{r-\alpha\,\ell}{M}}\right)
\ .
\end{eqnarray}
The event horizon is now at $r_{\rm H}=\alpha\,\ell=\ell_0\geq 2\,M$.
If we further assume $\alpha\,\ell\sim\,M$, we can express $M=M(Q)$ from~\eqref{qcase2} 
and the metric functions~\eqref{dominantBH3} can again be interpreted as a
nonlinear electrodynamics coupled with gravity.  
\subsubsection*{Case 3}
Finally, we consider
\begin{equation}
Q^2
=
\alpha\,M\left(2\,M+\alpha\,\ell\right)
e^{-\frac{(2\,M+\alpha\,\ell)}{M}}
\ ,
\end{equation}
so that
\begin{eqnarray}
\label{dominantBH4}
e^{\nu}
=
e^{-\lambda}
&=&
1-\frac{2\,M+\alpha\,\ell}{r}
-\frac{\alpha\,M}{r^2}\,e^{-r/M}
\nonumber
\\
&&
\times\left[r-\left(2\,M+\alpha\,\ell\right)e^{\frac{r-\left(2\,M+\alpha\,\ell\right)}{M}}\right]
\ .
\end{eqnarray}
The event horizon is at $r_{\rm H}=2\,M+\alpha\,\ell=2\,{\cal M}\geq 2\,M$, and the Schwarzschild horizon
is recovered as usual for $\alpha=0$.
As in the two previous cases, the interpretation in terms of nonlinear electrodynamics
is obtained for $\alpha\,\ell\sim M$. 
\par
The metric functions for the three analytical cases in Eqs.~\eqref{dominantBH2},
\eqref{dominantBH3} and~\eqref{dominantBH4} are displayed in Fig.~\ref{fig3},
each for a different value of the parameter $\ell$.
In all cases the density and pressures have the same qualitative behavior as
in Fig.~\ref{fig4}, which corresponds to the general case given by
Eqs.~\eqref{dominantBH}-\eqref{predom}.
We conclude that the metric~\eqref{dominantBH} represents a hairy black hole
endowed with the parameters $M$, $Q$, $\alpha$ and $\ell$, where $\{Q,\ell_0=\alpha\,\ell\}$
represents a potential set of charges generating primary hair.
Among this new family of solutions, we can identify three different cases
representing hairy black holes having simple analytical horizons.
All these cases can be interpreted as a nonlinear electrodynamics
coupled with gravity, whose charges are $M$, $\ell_0=\alpha\,\ell$ and $Q$.~\footnote{This way
of presenting the results should not overshadow the fact that $Q^2\sim\alpha$, so that the
Schwarzschild geometry is always recovered for $\alpha\to 0$.}
\par
Finally, we want to end by emphasizing a rather important result.
When the horizon in~\eqref{dominantBHH} has the simple form $r_{\rm H}=K\,M$,
with $K\ge 2$ in order to satisfy the DEC, the metric functions become
\begin{equation}
\label{dominantBHf}
e^{\nu}
=
e^{-\lambda}
=
1-\frac{2\,{\cal M}}{r}
+\frac{Q^2}{r^2}
-\frac{\alpha\,r_{\rm H}}{K\,r}e^{-K\,r/r_{\rm H}}
\ ,
\end{equation}
and the event horizon takes the simple Reissner-Nordstr\"om  form
\begin{equation}
\label{domhorf}
r_{\rm H}
=
\bar{\cal M}
+\sqrt{\bar{\cal M}^2-\bar{Q}^2}
\ ,
\end{equation}
where $\bar{\cal M}={\cal M}/\beta$ and $\bar{Q}^2={Q}^2/\beta$ with
\begin{equation}
\beta=1-\alpha\,\frac{e^{-K}}{K}
\ .
\end{equation}
We conclude that the metric~\eqref{dominantBHf} represents a black hole associated
with some non-linear electrodynamics, whose horizon is related with the Reissner-Nordstr\"om
one by the quite simple expression
\begin{eqnarray}
r_{\rm H}
=
\frac{r_{\rm RN}}{\beta}
\geq
r_{\rm RN}
\ ,
\end{eqnarray}
since $\beta\leq 1$. In order to specify this nonlinear electrodynamics, we identify
\begin{eqnarray}
\label{tmunu}
\theta_{\mu\nu}
=
-\left[
L_{F}\,F_{\mu\alpha}\,F^{\alpha}{} _{\nu}
+L(F)\,g_{\mu\nu}
\right]
\ ,
\end{eqnarray}
where 
\begin{equation}
F
=
\frac{1}{4}\,F_{\mu\nu}\,F^{\mu\nu}
\quad
{\rm and}
\quad 
L_F=\frac{dL}{dF}
\ .
\end{equation}
In the static spherically symmetric case, we have
\begin{equation}
F_{\mu\nu}
=
E(r)\,\left(\delta^0_\mu\,\delta^1_\nu-\delta^1_\mu\,\delta^0_\nu\right)
\ .
\end{equation}
Following a standard procedure, we then obtain the electric field
\begin{eqnarray}
E(r)
=
\frac{Q}{r^2}
-\frac{\alpha\,  e^{-\frac{K\,r}{r_{\rm H}}} (K\,r+2\,r_{\rm H})}
{4\,r_{\rm H}\,Q}
\ .
\end{eqnarray}
We can describe the underlying nonlinear electrodynamics within the
$P$-dual formalism~\cite{Salazar:1987ap,AyonBeato:1998ub}, which yields
the Lagrangian 
\begin{eqnarray}
\label{Lp}
L(P)
=
-4\, \pi\,  P
-\frac{\alpha\,K\,\sqrt[4]{-2\,P}\,e^{\mathcal{G}(P)}}
{4\, \sqrt{\pi }\, \sqrt{Q}\, r_{\rm H}}
\ ,
\end{eqnarray}
with
\begin{eqnarray}
\mathcal{G}(P)
=
-\frac{K\,\sqrt{Q}}{2\, \sqrt{\pi }\, \sqrt[4]{-2\,P}\,r_{\rm H}}
\ ,
\end{eqnarray}
where 
\begin{equation}
P
=
\frac{1}{4}\,P_{\mu\nu}\,P^{\mu\nu}
\quad
{\rm and}
\quad 
P_{\mu\nu}=L_F\,F_{\mu\nu}
\ .
\end{equation}
\par
We want to conclude emphasizing once again that $Q^2\sim\alpha$,
so that the Schwarzschild geometry is always recovered for $\alpha\to 0$.
However, when $Q$ represents an electric charge, we can say that the
electro-vacuum (Reissner-Nordstr\"om geometry) is filled with a tensor-vacuum
whose origin lies in the nonlinear electrodynamics with the Lagrangian~\eqref{Lp}. 
\section{Conclusions}
\label{Scon}
Using the EGD approach, we studied the emergence of hairy black holes due to matter
surrounding the central source of the Schwarzschild metric.
Demanding that the solution always admits a well defined horizon through Eq.~\eqref{constr1},
and that the hair satisfies the SEC or DEC through Eqs.~\eqref{strong01} or~\eqref{dom1}
and~\eqref{dom2}, respectively, we found two new families of hairy black holes displayed
in Eqs.~\eqref{strongBH} and~\eqref{dominantBH}.
These geometries were analysed and, in particular, we found that the solutions
satisfying the DEC contain the non-trivial extension of the Reissner-Nordstr\"om
black hole shown in~\eqref{dominantBHf} which possesses a simple event horizon.
The Lagrangian of the nonlinear electrodynamics, which sources this solution, was also
obtained explicitly.
\par
From the technical point of view, since aim of the present work was to study some general
conditions under which hair can be added to the (spherically symmetric) vacuum
black holes of general relativity, the EGD was the natural approach to employ from the onset.
In fact, the EGD is precisely devised for describing deformations of known solutions
of general relativity induced by adding extra sources.
Moreover, the properties of the added source were restricted in order to enforce
the conditions mentioned above, rather than assuming the hair is described in terms
of fundamental fields.
Nonetheless, we were able to give a description in terms of a nonlinear electrodynamics
at least in one specific case.
\par  
Finally, we would like to remark that the charges $Q$ and $\ell_0=\alpha\,\ell$
associated with our hairy black holes admit simple physical interpretations.
The charge $Q$ can be viewed as an effective electric charge and is proportional
to $\alpha$, which is the generic parameter measuring the deviation from the chosen
vacuum solution (which is given by the Schwarzschild metric).
A special mention deserves the parameter $\ell_0$, which is associated with gauge
transformations of the seed Schwarzschild metric and which seems to always push
the event horizon to radii larger than the Schwarzschild radius.
Therefore $\ell_0=\alpha\,\ell$ measures how much the entropy of the black hole increases from
its minimum Schwarzschild value $S=4\,\pi\,{M^2}$ when hair is added.
\subsection*{Acknowledgments}
R.C.~is partially supported by the INFN grant FLAG and his work has also been carried out in the framework
of activities of the National Group of Mathematical Physics (GNFM, INdAM) and COST action {\em Cantata\/}. 
%
%
%
\bibliography{references.bib}

\begin{thebibliography}{98}%
\makeatletter
\providecommand \@ifxundefined [1]{%
 \@ifx{#1\undefined}
}%
\providecommand \@ifnum [1]{%
 \ifnum #1\expandafter \@firstoftwo
 \else \expandafter \@secondoftwo
 \fi
}%
\providecommand \@ifx [1]{%
 \ifx #1\expandafter \@firstoftwo
 \else \expandafter \@secondoftwo
 \fi
}%
\providecommand \natexlab [1]{#1}%
\providecommand \enquote  [1]{``#1''}%
\providecommand \bibnamefont  [1]{#1}%
\providecommand \bibfnamefont [1]{#1}%
\providecommand \citenamefont [1]{#1}%
\providecommand \href@noop [0]{\@secondoftwo}%
\providecommand \href [0]{\begingroup \@sanitize@url \@href}%
\providecommand \@href[1]{\@@startlink{#1}\@@href}%
\providecommand \@@href[1]{\endgroup#1\@@endlink}%
\providecommand \@sanitize@url [0]{\catcode `\\12\catcode `\$12\catcode
  `\&12\catcode `\#12\catcode `\^12\catcode `\_12\catcode `\%12\relax}%
\providecommand \@@startlink[1]{}%
\providecommand \@@endlink[0]{}%
\providecommand \url  [0]{\begingroup\@sanitize@url \@url }%
\providecommand \@url [1]{\endgroup\@href {#1}{\urlprefix }}%
\providecommand \urlprefix  [0]{URL }%
\providecommand \Eprint [0]{\href }%
\providecommand \doibase [0]{http://dx.doi.org/}%
\providecommand \selectlanguage [0]{\@gobble}%
\providecommand \bibinfo  [0]{\@secondoftwo}%
\providecommand \bibfield  [0]{\@secondoftwo}%
\providecommand \translation [1]{[#1]}%
\providecommand \BibitemOpen [0]{}%
\providecommand \bibitemStop [0]{}%
\providecommand \bibitemNoStop [0]{.\EOS\space}%
\providecommand \EOS [0]{\spacefactor3000\relax}%
\providecommand \BibitemShut  [1]{\csname bibitem#1\endcsname}%
\let\auto@bib@innerbib\@empty
\bibitem [{\citenamefont {Abbott}\ \emph {et~al.}(2016)\citenamefont {Abbott}
  \emph {et~al.}}]{Abbott:2016blz}%
  \BibitemOpen
  \bibfield  {author} {\bibinfo {author} {\bibfnamefont {B.~P.}\ \bibnamefont
  {Abbott}} \emph {et~al.} (\bibinfo {collaboration} {LIGO Scientific,
  Virgo}),\ }\href {\doibase 10.1103/PhysRevLett.116.061102} {\bibfield
  {journal} {\bibinfo  {journal} {Phys. Rev. Lett.}\ }\textbf {\bibinfo
  {volume} {116}},\ \bibinfo {pages} {061102} (\bibinfo {year} {2016})},\
  \Eprint {http://arxiv.org/abs/1602.03837} {arXiv:1602.03837 [gr-qc]}
  \BibitemShut {NoStop}%
\bibitem [{\citenamefont {Abbott}\ \emph {et~al.}(2017)\citenamefont {Abbott}
  \emph {et~al.}}]{Abbott:2017oio}%
  \BibitemOpen
  \bibfield  {author} {\bibinfo {author} {\bibfnamefont {B.~P.}\ \bibnamefont
  {Abbott}} \emph {et~al.} (\bibinfo {collaboration} {LIGO Scientific,
  Virgo}),\ }\href {\doibase 10.1103/PhysRevLett.119.141101} {\bibfield
  {journal} {\bibinfo  {journal} {Phys. Rev. Lett.}\ }\textbf {\bibinfo
  {volume} {119}},\ \bibinfo {pages} {141101} (\bibinfo {year} {2017})},\
  \Eprint {http://arxiv.org/abs/1709.09660} {arXiv:1709.09660 [gr-qc]}
  \BibitemShut {NoStop}%
\bibitem [{\citenamefont {Akiyama}\ \emph {et~al.}(2019)\citenamefont {Akiyama}
  \emph {et~al.}}]{Akiyama:2019cqa}%
  \BibitemOpen
  \bibfield  {author} {\bibinfo {author} {\bibfnamefont {K.}~\bibnamefont
  {Akiyama}} \emph {et~al.} (\bibinfo {collaboration} {Event Horizon
  Telescope}),\ }\href {\doibase 10.3847/2041-8213/ab0ec7} {\bibfield
  {journal} {\bibinfo  {journal} {Astrophys. J.}\ }\textbf {\bibinfo {volume}
  {875}},\ \bibinfo {pages} {L1} (\bibinfo {year} {2019})}\BibitemShut
  {NoStop}%
\bibitem [{\citenamefont {Ruffini}\ and\ \citenamefont
  {Wheeler}(1971)}]{Ruffini:1971bza}%
  \BibitemOpen
  \bibfield  {author} {\bibinfo {author} {\bibfnamefont {R.}~\bibnamefont
  {Ruffini}}\ and\ \bibinfo {author} {\bibfnamefont {J.~A.}\ \bibnamefont
  {Wheeler}},\ }\href {\doibase 10.1063/1.3022513} {\bibfield  {journal}
  {\bibinfo  {journal} {Phys. Today}\ }\textbf {\bibinfo {volume} {24}},\
  \bibinfo {pages} {30} (\bibinfo {year} {1971})}\BibitemShut {NoStop}%
\bibitem [{\citenamefont {Hawking}(1972)}]{Hawking:1971vc}%
  \BibitemOpen
  \bibfield  {author} {\bibinfo {author} {\bibfnamefont {S.}~\bibnamefont
  {Hawking}},\ }\href {\doibase 10.1007/BF01877517} {\bibfield  {journal}
  {\bibinfo  {journal} {Commun. Math. Phys.}\ }\textbf {\bibinfo {volume}
  {25}},\ \bibinfo {pages} {152} (\bibinfo {year} {1972})}\BibitemShut
  {NoStop}%
\bibitem [{\citenamefont {Hawking}\ \emph {et~al.}(2016)\citenamefont
  {Hawking}, \citenamefont {Perry},\ and\ \citenamefont
  {Strominger}}]{Hawking:2016msc}%
  \BibitemOpen
  \bibfield  {author} {\bibinfo {author} {\bibfnamefont {S.~W.}\ \bibnamefont
  {Hawking}}, \bibinfo {author} {\bibfnamefont {M.~J.}\ \bibnamefont {Perry}},
  \ and\ \bibinfo {author} {\bibfnamefont {A.}~\bibnamefont {Strominger}},\
  }\href {\doibase 10.1103/PhysRevLett.116.231301} {\bibfield  {journal}
  {\bibinfo  {journal} {Phys. Rev. Lett.}\ }\textbf {\bibinfo {volume} {116}},\
  \bibinfo {pages} {231301} (\bibinfo {year} {2016})},\ \Eprint
  {http://arxiv.org/abs/1601.00921} {arXiv:1601.00921 [hep-th]} \BibitemShut
  {NoStop}%
\bibitem [{\citenamefont {Sotiriou}\ and\ \citenamefont
  {Faraoni}(2012)}]{Sotiriou:2011dz}%
  \BibitemOpen
  \bibfield  {author} {\bibinfo {author} {\bibfnamefont {T.~P.}\ \bibnamefont
  {Sotiriou}}\ and\ \bibinfo {author} {\bibfnamefont {V.}~\bibnamefont
  {Faraoni}},\ }\href {\doibase 10.1103/PhysRevLett.108.081103} {\bibfield
  {journal} {\bibinfo  {journal} {Phys. Rev. Lett.}\ }\textbf {\bibinfo
  {volume} {108}},\ \bibinfo {pages} {081103} (\bibinfo {year} {2012})},\
  \Eprint {http://arxiv.org/abs/1109.6324} {arXiv:1109.6324 [gr-qc]}
  \BibitemShut {NoStop}%
\bibitem [{\citenamefont {Babichev}\ and\ \citenamefont
  {Charmousis}(2014)}]{Babichev:2013cya}%
  \BibitemOpen
  \bibfield  {author} {\bibinfo {author} {\bibfnamefont {E.}~\bibnamefont
  {Babichev}}\ and\ \bibinfo {author} {\bibfnamefont {C.}~\bibnamefont
  {Charmousis}},\ }\href {\doibase 10.1007/JHEP08(2014)106} {\bibfield
  {journal} {\bibinfo  {journal} {JHEP}\ }\textbf {\bibinfo {volume} {08}},\
  \bibinfo {pages} {106} (\bibinfo {year} {2014})},\ \Eprint
  {http://arxiv.org/abs/1312.3204} {arXiv:1312.3204 [gr-qc]} \BibitemShut
  {NoStop}%
\bibitem [{\citenamefont {Cisterna}\ and\ \citenamefont
  {Erices}(2014)}]{Cisterna:2014nua}%
  \BibitemOpen
  \bibfield  {author} {\bibinfo {author} {\bibfnamefont {A.}~\bibnamefont
  {Cisterna}}\ and\ \bibinfo {author} {\bibfnamefont {C.}~\bibnamefont
  {Erices}},\ }\href {\doibase 10.1103/PhysRevD.89.084038} {\bibfield
  {journal} {\bibinfo  {journal} {Phys. Rev. D}\ }\textbf {\bibinfo {volume}
  {89}},\ \bibinfo {pages} {084038} (\bibinfo {year} {2014})},\ \Eprint
  {http://arxiv.org/abs/1401.4479} {arXiv:1401.4479 [gr-qc]} \BibitemShut
  {NoStop}%
\bibitem [{\citenamefont {Sotiriou}\ and\ \citenamefont
  {Zhou}(2014)}]{Sotiriou:2013qea}%
  \BibitemOpen
  \bibfield  {author} {\bibinfo {author} {\bibfnamefont {T.~P.}\ \bibnamefont
  {Sotiriou}}\ and\ \bibinfo {author} {\bibfnamefont {S.-Y.}\ \bibnamefont
  {Zhou}},\ }\href {\doibase 10.1103/PhysRevLett.112.251102} {\bibfield
  {journal} {\bibinfo  {journal} {Phys. Rev. Lett.}\ }\textbf {\bibinfo
  {volume} {112}},\ \bibinfo {pages} {251102} (\bibinfo {year} {2014})},\
  \Eprint {http://arxiv.org/abs/1312.3622} {arXiv:1312.3622 [gr-qc]}
  \BibitemShut {NoStop}%
\bibitem [{\citenamefont {Antoniou}\ \emph
  {et~al.}(2018{\natexlab{a}})\citenamefont {Antoniou}, \citenamefont
  {Bakopoulos},\ and\ \citenamefont {Kanti}}]{Antoniou:2017acq}%
  \BibitemOpen
  \bibfield  {author} {\bibinfo {author} {\bibfnamefont {G.}~\bibnamefont
  {Antoniou}}, \bibinfo {author} {\bibfnamefont {A.}~\bibnamefont
  {Bakopoulos}}, \ and\ \bibinfo {author} {\bibfnamefont {P.}~\bibnamefont
  {Kanti}},\ }\href {\doibase 10.1103/PhysRevLett.120.131102} {\bibfield
  {journal} {\bibinfo  {journal} {Phys. Rev. Lett.}\ }\textbf {\bibinfo
  {volume} {120}},\ \bibinfo {pages} {131102} (\bibinfo {year}
  {2018}{\natexlab{a}})},\ \Eprint {http://arxiv.org/abs/1711.03390}
  {arXiv:1711.03390 [hep-th]} \BibitemShut {NoStop}%
\bibitem [{\citenamefont {Antoniou}\ \emph
  {et~al.}(2018{\natexlab{b}})\citenamefont {Antoniou}, \citenamefont
  {Bakopoulos},\ and\ \citenamefont {Kanti}}]{Antoniou:2017hxj}%
  \BibitemOpen
  \bibfield  {author} {\bibinfo {author} {\bibfnamefont {G.}~\bibnamefont
  {Antoniou}}, \bibinfo {author} {\bibfnamefont {A.}~\bibnamefont
  {Bakopoulos}}, \ and\ \bibinfo {author} {\bibfnamefont {P.}~\bibnamefont
  {Kanti}},\ }\href {\doibase 10.1103/PhysRevD.97.084037} {\bibfield  {journal}
  {\bibinfo  {journal} {Phys. Rev. D}\ }\textbf {\bibinfo {volume} {97}},\
  \bibinfo {pages} {084037} (\bibinfo {year} {2018}{\natexlab{b}})},\ \Eprint
  {http://arxiv.org/abs/1711.07431} {arXiv:1711.07431 [hep-th]} \BibitemShut
  {NoStop}%
\bibitem [{\citenamefont {Grumiller}\ \emph {et~al.}(2020)\citenamefont
  {Grumiller}, \citenamefont {Pérez}, \citenamefont {Sheikh-Jabbari},
  \citenamefont {Troncoso},\ and\ \citenamefont {Zwikel}}]{Grumiller:2019fmp}%
  \BibitemOpen
  \bibfield  {author} {\bibinfo {author} {\bibfnamefont {D.}~\bibnamefont
  {Grumiller}}, \bibinfo {author} {\bibfnamefont {A.}~\bibnamefont {Pérez}},
  \bibinfo {author} {\bibfnamefont {M.}~\bibnamefont {Sheikh-Jabbari}},
  \bibinfo {author} {\bibfnamefont {R.}~\bibnamefont {Troncoso}}, \ and\
  \bibinfo {author} {\bibfnamefont {C.}~\bibnamefont {Zwikel}},\ }\href
  {\doibase 10.1103/PhysRevLett.124.041601} {\bibfield  {journal} {\bibinfo
  {journal} {Phys. Rev. Lett.}\ }\textbf {\bibinfo {volume} {124}},\ \bibinfo
  {pages} {041601} (\bibinfo {year} {2020})},\ \Eprint
  {http://arxiv.org/abs/1908.09833} {arXiv:1908.09833 [hep-th]} \BibitemShut
  {NoStop}%
\bibitem [{\citenamefont {Volkov}\ and\ \citenamefont
  {Galtsov}(1989)}]{Volkov:1989fi}%
  \BibitemOpen
  \bibfield  {author} {\bibinfo {author} {\bibfnamefont {M.}~\bibnamefont
  {Volkov}}\ and\ \bibinfo {author} {\bibfnamefont {D.}~\bibnamefont
  {Galtsov}},\ }\href@noop {} {\bibfield  {journal} {\bibinfo  {journal} {JETP
  Lett.}\ }\textbf {\bibinfo {volume} {50}},\ \bibinfo {pages} {346} (\bibinfo
  {year} {1989})}\BibitemShut {NoStop}%
\bibitem [{\citenamefont {Kanti}\ \emph {et~al.}(1996)\citenamefont {Kanti},
  \citenamefont {Mavromatos}, \citenamefont {Rizos}, \citenamefont {Tamvakis},\
  and\ \citenamefont {Winstanley}}]{Kanti:1995vq}%
  \BibitemOpen
  \bibfield  {author} {\bibinfo {author} {\bibfnamefont {P.}~\bibnamefont
  {Kanti}}, \bibinfo {author} {\bibfnamefont {N.}~\bibnamefont {Mavromatos}},
  \bibinfo {author} {\bibfnamefont {J.}~\bibnamefont {Rizos}}, \bibinfo
  {author} {\bibfnamefont {K.}~\bibnamefont {Tamvakis}}, \ and\ \bibinfo
  {author} {\bibfnamefont {E.}~\bibnamefont {Winstanley}},\ }\href {\doibase
  10.1103/PhysRevD.54.5049} {\bibfield  {journal} {\bibinfo  {journal} {Phys.
  Rev. D}\ }\textbf {\bibinfo {volume} {54}},\ \bibinfo {pages} {5049}
  (\bibinfo {year} {1996})},\ \Eprint {http://arxiv.org/abs/hep-th/9511071}
  {arXiv:hep-th/9511071} \BibitemShut {NoStop}%
\bibitem [{\citenamefont {Kanti}\ \emph {et~al.}(1998)\citenamefont {Kanti},
  \citenamefont {Mavromatos}, \citenamefont {Rizos}, \citenamefont {Tamvakis},\
  and\ \citenamefont {Winstanley}}]{Kanti:1997br}%
  \BibitemOpen
  \bibfield  {author} {\bibinfo {author} {\bibfnamefont {P.}~\bibnamefont
  {Kanti}}, \bibinfo {author} {\bibfnamefont {N.}~\bibnamefont {Mavromatos}},
  \bibinfo {author} {\bibfnamefont {J.}~\bibnamefont {Rizos}}, \bibinfo
  {author} {\bibfnamefont {K.}~\bibnamefont {Tamvakis}}, \ and\ \bibinfo
  {author} {\bibfnamefont {E.}~\bibnamefont {Winstanley}},\ }\href {\doibase
  10.1103/PhysRevD.57.6255} {\bibfield  {journal} {\bibinfo  {journal} {Phys.
  Rev. D}\ }\textbf {\bibinfo {volume} {57}},\ \bibinfo {pages} {6255}
  (\bibinfo {year} {1998})},\ \Eprint {http://arxiv.org/abs/hep-th/9703192}
  {arXiv:hep-th/9703192} \BibitemShut {NoStop}%
\bibitem [{\citenamefont {Zloshchastiev}(2005)}]{Zloshchastiev:2004ny}%
  \BibitemOpen
  \bibfield  {author} {\bibinfo {author} {\bibfnamefont {K.~G.}\ \bibnamefont
  {Zloshchastiev}},\ }\href {\doibase 10.1103/PhysRevLett.94.121101} {\bibfield
   {journal} {\bibinfo  {journal} {Phys. Rev. Lett.}\ }\textbf {\bibinfo
  {volume} {94}},\ \bibinfo {pages} {121101} (\bibinfo {year} {2005})},\
  \Eprint {http://arxiv.org/abs/hep-th/0408163} {arXiv:hep-th/0408163}
  \BibitemShut {NoStop}%
\bibitem [{\citenamefont {Martinez}\ \emph {et~al.}(2004)\citenamefont
  {Martinez}, \citenamefont {Troncoso},\ and\ \citenamefont
  {Zanelli}}]{Martinez:2004nb}%
  \BibitemOpen
  \bibfield  {author} {\bibinfo {author} {\bibfnamefont {C.}~\bibnamefont
  {Martinez}}, \bibinfo {author} {\bibfnamefont {R.}~\bibnamefont {Troncoso}},
  \ and\ \bibinfo {author} {\bibfnamefont {J.}~\bibnamefont {Zanelli}},\ }\href
  {\doibase 10.1103/PhysRevD.70.084035} {\bibfield  {journal} {\bibinfo
  {journal} {Phys. Rev. D}\ }\textbf {\bibinfo {volume} {70}},\ \bibinfo
  {pages} {084035} (\bibinfo {year} {2004})},\ \Eprint
  {http://arxiv.org/abs/hep-th/0406111} {arXiv:hep-th/0406111} \BibitemShut
  {NoStop}%
\bibitem [{\citenamefont {Herdeiro}\ and\ \citenamefont
  {Radu}(2015)}]{Herdeiro:2015waa}%
  \BibitemOpen
  \bibfield  {author} {\bibinfo {author} {\bibfnamefont {C.~A.}\ \bibnamefont
  {Herdeiro}}\ and\ \bibinfo {author} {\bibfnamefont {E.}~\bibnamefont
  {Radu}},\ }\href {\doibase 10.1142/S0218271815420146} {\bibfield  {journal}
  {\bibinfo  {journal} {Int. J. Mod. Phys. D}\ }\textbf {\bibinfo {volume}
  {24}},\ \bibinfo {pages} {1542014} (\bibinfo {year} {2015})},\ \Eprint
  {http://arxiv.org/abs/1504.08209} {arXiv:1504.08209 [gr-qc]} \BibitemShut
  {NoStop}%
\bibitem [{\citenamefont {Sotiriou}(2015)}]{Sotiriou:2015pka}%
  \BibitemOpen
  \bibfield  {author} {\bibinfo {author} {\bibfnamefont {T.~P.}\ \bibnamefont
  {Sotiriou}},\ }\href {\doibase 10.1088/0264-9381/32/21/214002} {\bibfield
  {journal} {\bibinfo  {journal} {Class. Quant. Grav.}\ }\textbf {\bibinfo
  {volume} {32}},\ \bibinfo {pages} {214002} (\bibinfo {year} {2015})},\
  \Eprint {http://arxiv.org/abs/1505.00248} {arXiv:1505.00248 [gr-qc]}
  \BibitemShut {NoStop}%
\bibitem [{\citenamefont {Ovalle}\ \emph
  {et~al.}(2018{\natexlab{a}})\citenamefont {Ovalle}, \citenamefont {Casadio},
  \citenamefont {Rocha}, \citenamefont {Sotomayor},\ and\ \citenamefont
  {Stuchlik}}]{Ovalle:2018umz}%
  \BibitemOpen
  \bibfield  {author} {\bibinfo {author} {\bibfnamefont {J.}~\bibnamefont
  {Ovalle}}, \bibinfo {author} {\bibfnamefont {R.}~\bibnamefont {Casadio}},
  \bibinfo {author} {\bibfnamefont {R.~d.}\ \bibnamefont {Rocha}}, \bibinfo
  {author} {\bibfnamefont {A.}~\bibnamefont {Sotomayor}}, \ and\ \bibinfo
  {author} {\bibfnamefont {Z.}~\bibnamefont {Stuchlik}},\ }\href {\doibase
  10.1140/epjc/s10052-018-6450-4} {\bibfield  {journal} {\bibinfo  {journal}
  {Eur. Phys. J.}\ }\textbf {\bibinfo {volume} {C78}},\ \bibinfo {pages} {960}
  (\bibinfo {year} {2018}{\natexlab{a}})},\ \Eprint
  {http://arxiv.org/abs/1804.03468} {arXiv:1804.03468 [gr-qc]} \BibitemShut
  {NoStop}%
\bibitem [{\citenamefont {Ovalle}(2019)}]{Ovalle:2019qyi}%
  \BibitemOpen
  \bibfield  {author} {\bibinfo {author} {\bibfnamefont {J.}~\bibnamefont
  {Ovalle}},\ }\href {\doibase 10.1016/j.physletb.2018.11.029} {\bibfield
  {journal} {\bibinfo  {journal} {Phys. Lett.}\ }\textbf {\bibinfo {volume}
  {B788}},\ \bibinfo {pages} {213} (\bibinfo {year} {2019})},\ \Eprint
  {http://arxiv.org/abs/1812.03000} {arXiv:1812.03000 [gr-qc]} \BibitemShut
  {NoStop}%
\bibitem [{\citenamefont {Ovalle}(2017)}]{Ovalle:2017fgl}%
  \BibitemOpen
  \bibfield  {author} {\bibinfo {author} {\bibfnamefont {J.}~\bibnamefont
  {Ovalle}},\ }\href {\doibase 10.1103/PhysRevD.95.104019} {\bibfield
  {journal} {\bibinfo  {journal} {Phys. Rev.}\ }\textbf {\bibinfo {volume}
  {D95}},\ \bibinfo {pages} {104019} (\bibinfo {year} {2017})},\ \Eprint
  {http://arxiv.org/abs/1704.05899} {arXiv:1704.05899 [gr-qc]} \BibitemShut
  {NoStop}%
\bibitem [{\citenamefont {Ovalle}(2008)}]{Ovalle:2007bn}%
  \BibitemOpen
  \bibfield  {author} {\bibinfo {author} {\bibfnamefont {J.}~\bibnamefont
  {Ovalle}},\ }\href {\doibase 10.1142/S0217732308027011} {\bibfield  {journal}
  {\bibinfo  {journal} {Mod. Phys. Lett.}\ }\textbf {\bibinfo {volume} {A23}},\
  \bibinfo {pages} {3247} (\bibinfo {year} {2008})},\ \Eprint
  {http://arxiv.org/abs/gr-qc/0703095} {arXiv:gr-qc/0703095 [gr-qc]}
  \BibitemShut {NoStop}%
\bibitem [{\citenamefont {Ovalle}\ and\ \citenamefont
  {Casadio}(2020)}]{Ovalle:2020fuo}%
  \BibitemOpen
  \bibfield  {author} {\bibinfo {author} {\bibfnamefont {J.}~\bibnamefont
  {Ovalle}}\ and\ \bibinfo {author} {\bibfnamefont {R.}~\bibnamefont
  {Casadio}},\ }\href {\doibase 10.1007/978-3-030-39493-6} {\emph {\bibinfo
  {title} {{Beyond Einstein Gravity}}}},\ SpringerBriefs in Physics\ (\bibinfo
  {publisher} {Springer Nature},\ \bibinfo {address} {Cham},\ \bibinfo {year}
  {2020})\BibitemShut {NoStop}%
\bibitem [{\citenamefont {Ovalle}(2009)}]{Ovalle:2008se}%
  \BibitemOpen
  \bibfield  {author} {\bibinfo {author} {\bibfnamefont {J.}~\bibnamefont
  {Ovalle}},\ }\href {\doibase 10.1142/S0218271809014790} {\bibfield  {journal}
  {\bibinfo  {journal} {Int. J. Mod. Phys. D}\ }\textbf {\bibinfo {volume}
  {18}},\ \bibinfo {pages} {837} (\bibinfo {year} {2009})},\ \Eprint
  {http://arxiv.org/abs/0809.3547} {arXiv:0809.3547 [gr-qc]} \BibitemShut
  {NoStop}%
\bibitem [{\citenamefont {Ovalle}(2010)}]{Ovalle:2010zc}%
  \BibitemOpen
  \bibfield  {author} {\bibinfo {author} {\bibfnamefont {J.}~\bibnamefont
  {Ovalle}},\ }\href {\doibase 10.1142/S0217732310034420} {\bibfield  {journal}
  {\bibinfo  {journal} {Mod. Phys. Lett. A}\ }\textbf {\bibinfo {volume}
  {25}},\ \bibinfo {pages} {3323} (\bibinfo {year} {2010})},\ \Eprint
  {http://arxiv.org/abs/1009.3674} {arXiv:1009.3674 [gr-qc]} \BibitemShut
  {NoStop}%
\bibitem [{\citenamefont {Casadio}\ and\ \citenamefont
  {Ovalle}(2012)}]{Casadio:2012pu}%
  \BibitemOpen
  \bibfield  {author} {\bibinfo {author} {\bibfnamefont {R.}~\bibnamefont
  {Casadio}}\ and\ \bibinfo {author} {\bibfnamefont {J.}~\bibnamefont
  {Ovalle}},\ }\href {\doibase 10.1016/j.physletb.2012.07.041} {\bibfield
  {journal} {\bibinfo  {journal} {Phys. Lett.}\ }\textbf {\bibinfo {volume}
  {B715}},\ \bibinfo {pages} {251} (\bibinfo {year} {2012})},\ \Eprint
  {http://arxiv.org/abs/1201.6145} {arXiv:1201.6145 [gr-qc]} \BibitemShut
  {NoStop}%
\bibitem [{\citenamefont {Casadio}\ and\ \citenamefont
  {Ovalle}(2014)}]{Casadio:2012rf}%
  \BibitemOpen
  \bibfield  {author} {\bibinfo {author} {\bibfnamefont {R.}~\bibnamefont
  {Casadio}}\ and\ \bibinfo {author} {\bibfnamefont {J.}~\bibnamefont
  {Ovalle}},\ }\href {\doibase 10.1007/s10714-014-1669-3} {\bibfield  {journal}
  {\bibinfo  {journal} {Gen. Rel. Grav.}\ }\textbf {\bibinfo {volume} {46}},\
  \bibinfo {pages} {1669} (\bibinfo {year} {2014})},\ \Eprint
  {http://arxiv.org/abs/1212.0409} {arXiv:1212.0409 [gr-qc]} \BibitemShut
  {NoStop}%
\bibitem [{\citenamefont {Ovalle}\ \emph {et~al.}(2013)\citenamefont {Ovalle},
  \citenamefont {Linares}, \citenamefont {Pasqua},\ and\ \citenamefont
  {Sotomayor}}]{Ovalle:2013vna}%
  \BibitemOpen
  \bibfield  {author} {\bibinfo {author} {\bibfnamefont {J.}~\bibnamefont
  {Ovalle}}, \bibinfo {author} {\bibfnamefont {F.}~\bibnamefont {Linares}},
  \bibinfo {author} {\bibfnamefont {A.}~\bibnamefont {Pasqua}}, \ and\ \bibinfo
  {author} {\bibfnamefont {A.}~\bibnamefont {Sotomayor}},\ }\href {\doibase
  10.1088/0264-9381/30/17/175019} {\bibfield  {journal} {\bibinfo  {journal}
  {Class. Quant. Grav.}\ }\textbf {\bibinfo {volume} {30}},\ \bibinfo {pages}
  {175019} (\bibinfo {year} {2013})},\ \Eprint {http://arxiv.org/abs/1304.5995}
  {arXiv:1304.5995 [gr-qc]} \BibitemShut {NoStop}%
\bibitem [{\citenamefont {Casadio}\ \emph {et~al.}(2014)\citenamefont
  {Casadio}, \citenamefont {Ovalle},\ and\ \citenamefont
  {da~Rocha}}]{Casadio:2013uma}%
  \BibitemOpen
  \bibfield  {author} {\bibinfo {author} {\bibfnamefont {R.}~\bibnamefont
  {Casadio}}, \bibinfo {author} {\bibfnamefont {J.}~\bibnamefont {Ovalle}}, \
  and\ \bibinfo {author} {\bibfnamefont {R.}~\bibnamefont {da~Rocha}},\ }\href
  {\doibase 10.1088/0264-9381/31/4/045016} {\bibfield  {journal} {\bibinfo
  {journal} {Class. Quant. Grav.}\ }\textbf {\bibinfo {volume} {31}},\ \bibinfo
  {pages} {045016} (\bibinfo {year} {2014})},\ \Eprint
  {http://arxiv.org/abs/1310.5853} {arXiv:1310.5853 [gr-qc]} \BibitemShut
  {NoStop}%
\bibitem [{\citenamefont {Ovalle}\ and\ \citenamefont
  {Linares}(2013)}]{Ovalle:2013xla}%
  \BibitemOpen
  \bibfield  {author} {\bibinfo {author} {\bibfnamefont {J.}~\bibnamefont
  {Ovalle}}\ and\ \bibinfo {author} {\bibfnamefont {F.}~\bibnamefont
  {Linares}},\ }\href {\doibase 10.1103/PhysRevD.88.104026} {\bibfield
  {journal} {\bibinfo  {journal} {Phys. Rev.}\ }\textbf {\bibinfo {volume}
  {D88}},\ \bibinfo {pages} {104026} (\bibinfo {year} {2013})},\ \Eprint
  {http://arxiv.org/abs/1311.1844} {arXiv:1311.1844 [gr-qc]} \BibitemShut
  {NoStop}%
\bibitem [{\citenamefont {Ovalle}\ \emph {et~al.}(2015)\citenamefont {Ovalle},
  \citenamefont {Gergely},\ and\ \citenamefont {Casadio}}]{Ovalle:2014uwa}%
  \BibitemOpen
  \bibfield  {author} {\bibinfo {author} {\bibfnamefont {J.}~\bibnamefont
  {Ovalle}}, \bibinfo {author} {\bibfnamefont {L.~A.}\ \bibnamefont {Gergely}},
  \ and\ \bibinfo {author} {\bibfnamefont {R.}~\bibnamefont {Casadio}},\ }\href
  {\doibase 10.1088/0264-9381/32/4/045015} {\bibfield  {journal} {\bibinfo
  {journal} {Class. Quant. Grav.}\ }\textbf {\bibinfo {volume} {32}},\ \bibinfo
  {pages} {045015} (\bibinfo {year} {2015})},\ \Eprint
  {http://arxiv.org/abs/1405.0252} {arXiv:1405.0252 [gr-qc]} \BibitemShut
  {NoStop}%
\bibitem [{\citenamefont {Casadio}\ \emph
  {et~al.}(2015{\natexlab{a}})\citenamefont {Casadio}, \citenamefont {Ovalle},\
  and\ \citenamefont {da~Rocha}}]{Casadio:2015jva}%
  \BibitemOpen
  \bibfield  {author} {\bibinfo {author} {\bibfnamefont {R.}~\bibnamefont
  {Casadio}}, \bibinfo {author} {\bibfnamefont {J.}~\bibnamefont {Ovalle}}, \
  and\ \bibinfo {author} {\bibfnamefont {R.}~\bibnamefont {da~Rocha}},\ }\href
  {\doibase 10.1209/0295-5075/110/40003} {\bibfield  {journal} {\bibinfo
  {journal} {EPL}\ }\textbf {\bibinfo {volume} {110}},\ \bibinfo {pages}
  {40003} (\bibinfo {year} {2015}{\natexlab{a}})},\ \Eprint
  {http://arxiv.org/abs/1503.02316} {arXiv:1503.02316 [gr-qc]} \BibitemShut
  {NoStop}%
\bibitem [{\citenamefont {Casadio}\ \emph
  {et~al.}(2015{\natexlab{b}})\citenamefont {Casadio}, \citenamefont {Ovalle},\
  and\ \citenamefont {da~Rocha}}]{Casadio:2015gea}%
  \BibitemOpen
  \bibfield  {author} {\bibinfo {author} {\bibfnamefont {R.}~\bibnamefont
  {Casadio}}, \bibinfo {author} {\bibfnamefont {J.}~\bibnamefont {Ovalle}}, \
  and\ \bibinfo {author} {\bibfnamefont {R.}~\bibnamefont {da~Rocha}},\ }\href
  {\doibase 10.1088/0264-9381/32/21/215020} {\bibfield  {journal} {\bibinfo
  {journal} {Class. Quant. Grav.}\ }\textbf {\bibinfo {volume} {32}},\ \bibinfo
  {pages} {215020} (\bibinfo {year} {2015}{\natexlab{b}})},\ \Eprint
  {http://arxiv.org/abs/1503.02873} {arXiv:1503.02873 [gr-qc]} \BibitemShut
  {NoStop}%
\bibitem [{\citenamefont {Ovalle}(2015)}]{Ovalle:2015nfa}%
  \BibitemOpen
  \bibfield  {author} {\bibinfo {author} {\bibfnamefont {J.}~\bibnamefont
  {Ovalle}},\ }\bibfield  {booktitle} {\emph {\bibinfo {booktitle}
  {{Proceedings, 9th Alexander Friedmann International Seminar on Gravitation
  and Cosmology and 3rd Satellite Symposium on the Casimir Effect: St.
  Petersburg, Russia, June 21-27, 2015}}},\ }\href {\doibase
  10.1142/S2010194516601320} {\  (\bibinfo {year} {2015}),\
  10.1142/S2010194516601320},\ \bibinfo {note} {[Int. J. Mod. Phys. Conf.
  Ser.41,1660132(2016)]},\ \Eprint {http://arxiv.org/abs/1510.00855}
  {arXiv:1510.00855 [gr-qc]} \BibitemShut {NoStop}%
\bibitem [{\citenamefont {Cavalcanti}\ \emph {et~al.}(2016)\citenamefont
  {Cavalcanti}, \citenamefont {da~Silva},\ and\ \citenamefont
  {da~Rocha}}]{Cavalcanti:2016mbe}%
  \BibitemOpen
  \bibfield  {author} {\bibinfo {author} {\bibfnamefont {R.~T.}\ \bibnamefont
  {Cavalcanti}}, \bibinfo {author} {\bibfnamefont {A.~G.}\ \bibnamefont
  {da~Silva}}, \ and\ \bibinfo {author} {\bibfnamefont {R.}~\bibnamefont
  {da~Rocha}},\ }\href {\doibase 10.1088/0264-9381/33/21/215007} {\bibfield
  {journal} {\bibinfo  {journal} {Class. Quant. Grav.}\ }\textbf {\bibinfo
  {volume} {33}},\ \bibinfo {pages} {215007} (\bibinfo {year} {2016})},\
  \Eprint {http://arxiv.org/abs/1605.01271} {arXiv:1605.01271 [gr-qc]}
  \BibitemShut {NoStop}%
\bibitem [{\citenamefont {da~Rocha}(2017{\natexlab{a}})}]{daRocha:2017cxu}%
  \BibitemOpen
  \bibfield  {author} {\bibinfo {author} {\bibfnamefont {R.}~\bibnamefont
  {da~Rocha}},\ }\href {\doibase 10.1103/PhysRevD.95.124017} {\bibfield
  {journal} {\bibinfo  {journal} {Phys. Rev.}\ }\textbf {\bibinfo {volume}
  {D95}},\ \bibinfo {pages} {124017} (\bibinfo {year} {2017}{\natexlab{a}})},\
  \Eprint {http://arxiv.org/abs/1701.00761} {arXiv:1701.00761 [hep-ph]}
  \BibitemShut {NoStop}%
\bibitem [{\citenamefont {da~Rocha}(2017{\natexlab{b}})}]{daRocha:2017lqj}%
  \BibitemOpen
  \bibfield  {author} {\bibinfo {author} {\bibfnamefont {R.}~\bibnamefont
  {da~Rocha}},\ }\href {\doibase 10.1140/epjc/s10052-017-4926-2} {\bibfield
  {journal} {\bibinfo  {journal} {Eur. Phys. J.}\ }\textbf {\bibinfo {volume}
  {C77}},\ \bibinfo {pages} {355} (\bibinfo {year} {2017}{\natexlab{b}})},\
  \Eprint {http://arxiv.org/abs/1703.01528} {arXiv:1703.01528 [hep-th]}
  \BibitemShut {NoStop}%
\bibitem [{\citenamefont {Fernandes-Silva}\ and\ \citenamefont
  {da~Rocha}(2018)}]{Fernandes-Silva:2017nec}%
  \BibitemOpen
  \bibfield  {author} {\bibinfo {author} {\bibfnamefont {A.}~\bibnamefont
  {Fernandes-Silva}}\ and\ \bibinfo {author} {\bibfnamefont {R.}~\bibnamefont
  {da~Rocha}},\ }\href {\doibase 10.1140/epjc/s10052-018-5754-8} {\bibfield
  {journal} {\bibinfo  {journal} {Eur. Phys. J.}\ }\textbf {\bibinfo {volume}
  {C78}},\ \bibinfo {pages} {271} (\bibinfo {year} {2018})},\ \Eprint
  {http://arxiv.org/abs/1708.08686} {arXiv:1708.08686 [hep-th]} \BibitemShut
  {NoStop}%
\bibitem [{\citenamefont {Casadio}\ \emph {et~al.}(2018)\citenamefont
  {Casadio}, \citenamefont {Nicolini},\ and\ \citenamefont
  {da~Rocha}}]{Casadio:2017sze}%
  \BibitemOpen
  \bibfield  {author} {\bibinfo {author} {\bibfnamefont {R.}~\bibnamefont
  {Casadio}}, \bibinfo {author} {\bibfnamefont {P.}~\bibnamefont {Nicolini}}, \
  and\ \bibinfo {author} {\bibfnamefont {R.}~\bibnamefont {da~Rocha}},\ }\href
  {\doibase 10.1088/1361-6382/aad664} {\bibfield  {journal} {\bibinfo
  {journal} {Class. Quant. Grav.}\ }\textbf {\bibinfo {volume} {35}},\ \bibinfo
  {pages} {185001} (\bibinfo {year} {2018})},\ \Eprint
  {http://arxiv.org/abs/1709.09704} {arXiv:1709.09704 [hep-th]} \BibitemShut
  {NoStop}%
\bibitem [{\citenamefont {Fernandes-Silva}\ \emph {et~al.}(2018)\citenamefont
  {Fernandes-Silva}, \citenamefont {Ferreira-Martins},\ and\ \citenamefont
  {Da~Rocha}}]{Fernandes-Silva:2018abr}%
  \BibitemOpen
  \bibfield  {author} {\bibinfo {author} {\bibfnamefont {A.}~\bibnamefont
  {Fernandes-Silva}}, \bibinfo {author} {\bibfnamefont {A.~J.}\ \bibnamefont
  {Ferreira-Martins}}, \ and\ \bibinfo {author} {\bibfnamefont
  {R.}~\bibnamefont {Da~Rocha}},\ }\href {\doibase
  10.1140/epjc/s10052-018-6123-3} {\bibfield  {journal} {\bibinfo  {journal}
  {Eur. Phys. J.}\ }\textbf {\bibinfo {volume} {C78}},\ \bibinfo {pages} {631}
  (\bibinfo {year} {2018})},\ \Eprint {http://arxiv.org/abs/1803.03336}
  {arXiv:1803.03336 [hep-th]} \BibitemShut {NoStop}%
\bibitem [{\citenamefont {Contreras}\ and\ \citenamefont
  {Bargue\~no}(2018)}]{Contreras:2018vph}%
  \BibitemOpen
  \bibfield  {author} {\bibinfo {author} {\bibfnamefont {E.}~\bibnamefont
  {Contreras}}\ and\ \bibinfo {author} {\bibfnamefont {P.}~\bibnamefont
  {Bargue\~no}},\ }\href {\doibase 10.1140/epjc/s10052-018-6048-x} {\bibfield
  {journal} {\bibinfo  {journal} {Eur. Phys. J.}\ }\textbf {\bibinfo {volume}
  {C78}},\ \bibinfo {pages} {558} (\bibinfo {year} {2018})},\ \Eprint
  {http://arxiv.org/abs/1805.10565} {arXiv:1805.10565 [gr-qc]} \BibitemShut
  {NoStop}%
\bibitem [{\citenamefont {Contreras}(2018)}]{Contreras:2018gzd}%
  \BibitemOpen
  \bibfield  {author} {\bibinfo {author} {\bibfnamefont {E.}~\bibnamefont
  {Contreras}},\ }\href {\doibase 10.1140/epjc/s10052-018-6168-3} {\bibfield
  {journal} {\bibinfo  {journal} {Eur. Phys. J.}\ }\textbf {\bibinfo {volume}
  {C78}},\ \bibinfo {pages} {678} (\bibinfo {year} {2018})},\ \Eprint
  {http://arxiv.org/abs/1807.03252} {arXiv:1807.03252 [gr-qc]} \BibitemShut
  {NoStop}%
\bibitem [{\citenamefont {Contreras}\ and\ \citenamefont
  {Bargueño}(2018)}]{Contreras:2018nfg}%
  \BibitemOpen
  \bibfield  {author} {\bibinfo {author} {\bibfnamefont {E.}~\bibnamefont
  {Contreras}}\ and\ \bibinfo {author} {\bibfnamefont {P.}~\bibnamefont
  {Bargueño}},\ }\href {\doibase 10.1140/epjc/s10052-018-6472-y} {\bibfield
  {journal} {\bibinfo  {journal} {Eur. Phys. J. C}\ }\textbf {\bibinfo {volume}
  {78}},\ \bibinfo {pages} {985} (\bibinfo {year} {2018})},\ \Eprint
  {http://arxiv.org/abs/1809.09820} {arXiv:1809.09820 [gr-qc]} \BibitemShut
  {NoStop}%
\bibitem [{\citenamefont {Panotopoulos}\ and\ \citenamefont
  {Rinc\'on}(2018)}]{Panotopoulos:2018law}%
  \BibitemOpen
  \bibfield  {author} {\bibinfo {author} {\bibfnamefont {G.}~\bibnamefont
  {Panotopoulos}}\ and\ \bibinfo {author} {\bibfnamefont {A.}~\bibnamefont
  {Rinc\'on}},\ }\href {\doibase 10.1140/epjc/s10052-018-6321-z} {\bibfield
  {journal} {\bibinfo  {journal} {Eur. Phys. J.}\ }\textbf {\bibinfo {volume}
  {C78}},\ \bibinfo {pages} {851} (\bibinfo {year} {2018})},\ \Eprint
  {http://arxiv.org/abs/1810.08830} {arXiv:1810.08830 [gr-qc]} \BibitemShut
  {NoStop}%
\bibitem [{\citenamefont {Da~Rocha}\ and\ \citenamefont
  {Tomaz}(2019)}]{daRocha:2019pla}%
  \BibitemOpen
  \bibfield  {author} {\bibinfo {author} {\bibfnamefont {R.}~\bibnamefont
  {Da~Rocha}}\ and\ \bibinfo {author} {\bibfnamefont {A.~A.}\ \bibnamefont
  {Tomaz}},\ }\href {\doibase 10.1140/epjc/s10052-019-7558-x} {\bibfield
  {journal} {\bibinfo  {journal} {Eur. Phys. J. C}\ }\textbf {\bibinfo {volume}
  {79}},\ \bibinfo {pages} {1035} (\bibinfo {year} {2019})},\ \Eprint
  {http://arxiv.org/abs/1905.01548} {arXiv:1905.01548 [hep-th]} \BibitemShut
  {NoStop}%
\bibitem [{\citenamefont {Las~Heras}\ and\ \citenamefont
  {León}(2019)}]{Heras:2019ibr}%
  \BibitemOpen
  \bibfield  {author} {\bibinfo {author} {\bibfnamefont {C.}~\bibnamefont
  {Las~Heras}}\ and\ \bibinfo {author} {\bibfnamefont {P.}~\bibnamefont
  {León}},\ }\href {\doibase 10.1140/epjc/s10052-019-7507-8} {\bibfield
  {journal} {\bibinfo  {journal} {Eur. Phys. J. C}\ }\textbf {\bibinfo {volume}
  {79}},\ \bibinfo {pages} {990} (\bibinfo {year} {2019})},\ \Eprint
  {http://arxiv.org/abs/1905.02380} {arXiv:1905.02380 [gr-qc]} \BibitemShut
  {NoStop}%
\bibitem [{\citenamefont {Rinc\'on}\ \emph {et~al.}(2019)\citenamefont
  {Rinc\'on}, \citenamefont {Gabbanelli}, \citenamefont {Contreras},\ and\
  \citenamefont {Tello-Ortiz}}]{Rincon:2019jal}%
  \BibitemOpen
  \bibfield  {author} {\bibinfo {author} {\bibfnamefont {A.}~\bibnamefont
  {Rinc\'on}}, \bibinfo {author} {\bibfnamefont {L.}~\bibnamefont
  {Gabbanelli}}, \bibinfo {author} {\bibfnamefont {E.}~\bibnamefont
  {Contreras}}, \ and\ \bibinfo {author} {\bibfnamefont {F.}~\bibnamefont
  {Tello-Ortiz}},\ }\href {\doibase 10.1140/epjc/s10052-019-7397-9} {\bibfield
  {journal} {\bibinfo  {journal} {Eur. Phys. J. C}\ }\textbf {\bibinfo {volume}
  {79}},\ \bibinfo {pages} {873} (\bibinfo {year} {2019})},\ \Eprint
  {http://arxiv.org/abs/1909.00500} {arXiv:1909.00500 [gr-qc]} \BibitemShut
  {NoStop}%
\bibitem [{\citenamefont {da~Rocha}(2020{\natexlab{a}})}]{daRocha:2020rda}%
  \BibitemOpen
  \bibfield  {author} {\bibinfo {author} {\bibfnamefont {R.}~\bibnamefont
  {da~Rocha}},\ }\href {\doibase 10.3390/sym12040508} {\bibfield  {journal}
  {\bibinfo  {journal} {Symmetry}\ }\textbf {\bibinfo {volume} {12}},\ \bibinfo
  {pages} {508} (\bibinfo {year} {2020}{\natexlab{a}})},\ \Eprint
  {http://arxiv.org/abs/2002.10972} {arXiv:2002.10972 [hep-th]} \BibitemShut
  {NoStop}%
\bibitem [{\citenamefont {Contreras}\ \emph {et~al.}(2020)\citenamefont
  {Contreras}, \citenamefont {Tello-Ortíz},\ and\ \citenamefont
  {Maurya}}]{Contreras:2020fcj}%
  \BibitemOpen
  \bibfield  {author} {\bibinfo {author} {\bibfnamefont {E.}~\bibnamefont
  {Contreras}}, \bibinfo {author} {\bibfnamefont {F.}~\bibnamefont
  {Tello-Ortíz}}, \ and\ \bibinfo {author} {\bibfnamefont {S.}~\bibnamefont
  {Maurya}},\ }\href@noop {} {\  (\bibinfo {year} {2020})},\ \Eprint
  {http://arxiv.org/abs/2002.12444} {arXiv:2002.12444 [gr-qc]} \BibitemShut
  {NoStop}%
\bibitem [{\citenamefont {Arias}\ \emph {et~al.}(2020)\citenamefont {Arias},
  \citenamefont {Tello~Ortiz},\ and\ \citenamefont
  {Contreras}}]{Arias:2020hwz}%
  \BibitemOpen
  \bibfield  {author} {\bibinfo {author} {\bibfnamefont {C.}~\bibnamefont
  {Arias}}, \bibinfo {author} {\bibfnamefont {F.}~\bibnamefont {Tello~Ortiz}},
  \ and\ \bibinfo {author} {\bibfnamefont {E.}~\bibnamefont {Contreras}},\
  }\href {\doibase 10.1140/epjc/s10052-020-8042-3} {\bibfield  {journal}
  {\bibinfo  {journal} {Eur. Phys. J. C}\ }\textbf {\bibinfo {volume} {80}},\
  \bibinfo {pages} {463} (\bibinfo {year} {2020})},\ \Eprint
  {http://arxiv.org/abs/2003.00256} {arXiv:2003.00256 [gr-qc]} \BibitemShut
  {NoStop}%
\bibitem [{\citenamefont {da~Rocha}(2020{\natexlab{b}})}]{daRocha:2020jdj}%
  \BibitemOpen
  \bibfield  {author} {\bibinfo {author} {\bibfnamefont {R.}~\bibnamefont
  {da~Rocha}},\ }\href@noop {} {\  (\bibinfo {year} {2020}{\natexlab{b}})},\
  \Eprint {http://arxiv.org/abs/2003.12852} {arXiv:2003.12852 [hep-th]}
  \BibitemShut {NoStop}%
\bibitem [{\citenamefont {Tello-Ortiz}\ \emph {et~al.}(2020)\citenamefont
  {Tello-Ortiz}, \citenamefont {Maurya},\ and\ \citenamefont
  {Gomez-Leyton}}]{Tello-Ortiz:2020euy}%
  \BibitemOpen
  \bibfield  {author} {\bibinfo {author} {\bibfnamefont {F.}~\bibnamefont
  {Tello-Ortiz}}, \bibinfo {author} {\bibfnamefont {S.}~\bibnamefont {Maurya}},
  \ and\ \bibinfo {author} {\bibfnamefont {Y.}~\bibnamefont {Gomez-Leyton}},\
  }\href {\doibase 10.1140/epjc/s10052-020-7882-1} {\bibfield  {journal}
  {\bibinfo  {journal} {Eur. Phys. J. C}\ }\textbf {\bibinfo {volume} {80}},\
  \bibinfo {pages} {324} (\bibinfo {year} {2020})}\BibitemShut {NoStop}%
\bibitem [{\citenamefont {da~Rocha}\ and\ \citenamefont
  {Tomaz}(2020)}]{daRocha:2020gee}%
  \BibitemOpen
  \bibfield  {author} {\bibinfo {author} {\bibfnamefont {R.}~\bibnamefont
  {da~Rocha}}\ and\ \bibinfo {author} {\bibfnamefont {A.~A.}\ \bibnamefont
  {Tomaz}},\ }\href@noop {} {\  (\bibinfo {year} {2020})},\ \Eprint
  {http://arxiv.org/abs/2005.02980} {arXiv:2005.02980 [hep-th]} \BibitemShut
  {NoStop}%
\bibitem [{\citenamefont {Meert}\ and\ \citenamefont
  {da~Rocha}(2020)}]{Meert:2020sqv}%
  \BibitemOpen
  \bibfield  {author} {\bibinfo {author} {\bibfnamefont {P.}~\bibnamefont
  {Meert}}\ and\ \bibinfo {author} {\bibfnamefont {R.}~\bibnamefont
  {da~Rocha}},\ }\href@noop {} {\  (\bibinfo {year} {2020})},\ \Eprint
  {http://arxiv.org/abs/2006.02564} {arXiv:2006.02564 [gr-qc]} \BibitemShut
  {NoStop}%
\bibitem [{\citenamefont {Ovalle}\ \emph
  {et~al.}(2018{\natexlab{b}})\citenamefont {Ovalle}, \citenamefont {Casadio},
  \citenamefont {da~Rocha},\ and\ \citenamefont {Sotomayor}}]{Ovalle:2017wqi}%
  \BibitemOpen
  \bibfield  {author} {\bibinfo {author} {\bibfnamefont {J.}~\bibnamefont
  {Ovalle}}, \bibinfo {author} {\bibfnamefont {R.}~\bibnamefont {Casadio}},
  \bibinfo {author} {\bibfnamefont {R.}~\bibnamefont {da~Rocha}}, \ and\
  \bibinfo {author} {\bibfnamefont {A.}~\bibnamefont {Sotomayor}},\ }\href
  {\doibase 10.1140/epjc/s10052-018-5606-6} {\bibfield  {journal} {\bibinfo
  {journal} {Eur. Phys. J.}\ }\textbf {\bibinfo {volume} {C78}},\ \bibinfo
  {pages} {122} (\bibinfo {year} {2018}{\natexlab{b}})},\ \Eprint
  {http://arxiv.org/abs/1708.00407} {arXiv:1708.00407 [gr-qc]} \BibitemShut
  {NoStop}%
\bibitem [{\citenamefont {Gabbanelli}\ \emph {et~al.}(2018)\citenamefont
  {Gabbanelli}, \citenamefont {Rinc\'on},\ and\ \citenamefont
  {Rubio}}]{Gabbanelli:2018bhs}%
  \BibitemOpen
  \bibfield  {author} {\bibinfo {author} {\bibfnamefont {L.}~\bibnamefont
  {Gabbanelli}}, \bibinfo {author} {\bibfnamefont {A.}~\bibnamefont
  {Rinc\'on}}, \ and\ \bibinfo {author} {\bibfnamefont {C.}~\bibnamefont
  {Rubio}},\ }\href {\doibase 10.1140/epjc/s10052-018-5865-2} {\bibfield
  {journal} {\bibinfo  {journal} {Eur. Phys. J.}\ }\textbf {\bibinfo {volume}
  {C78}},\ \bibinfo {pages} {370} (\bibinfo {year} {2018})},\ \Eprint
  {http://arxiv.org/abs/1802.08000} {arXiv:1802.08000 [gr-qc]} \BibitemShut
  {NoStop}%
\bibitem [{\citenamefont {Heras}\ and\ \citenamefont
  {Leon}(2018)}]{Heras:2018cpz}%
  \BibitemOpen
  \bibfield  {author} {\bibinfo {author} {\bibfnamefont {C.~L.}\ \bibnamefont
  {Heras}}\ and\ \bibinfo {author} {\bibfnamefont {P.}~\bibnamefont {Leon}},\
  }\href {\doibase 10.1002/prop.201800036} {\bibfield  {journal} {\bibinfo
  {journal} {Fortsch. Phys.}\ }\textbf {\bibinfo {volume} {66}},\ \bibinfo
  {pages} {1800036} (\bibinfo {year} {2018})},\ \Eprint
  {http://arxiv.org/abs/1804.06874} {arXiv:1804.06874 [gr-qc]} \BibitemShut
  {NoStop}%
\bibitem [{\citenamefont {Estrada}\ and\ \citenamefont
  {Tello-Ortiz}(2018)}]{Estrada:2018zbh}%
  \BibitemOpen
  \bibfield  {author} {\bibinfo {author} {\bibfnamefont {M.}~\bibnamefont
  {Estrada}}\ and\ \bibinfo {author} {\bibfnamefont {F.}~\bibnamefont
  {Tello-Ortiz}},\ }\href {\doibase 10.1140/epjp/i2018-12249-9} {\bibfield
  {journal} {\bibinfo  {journal} {Eur. Phys. J. Plus}\ }\textbf {\bibinfo
  {volume} {133}},\ \bibinfo {pages} {453} (\bibinfo {year} {2018})},\ \Eprint
  {http://arxiv.org/abs/1803.02344} {arXiv:1803.02344 [gr-qc]} \BibitemShut
  {NoStop}%
\bibitem [{\citenamefont {Sharif}\ and\ \citenamefont
  {Sadiq}(2018{\natexlab{a}})}]{Sharif:2018toc}%
  \BibitemOpen
  \bibfield  {author} {\bibinfo {author} {\bibfnamefont {M.}~\bibnamefont
  {Sharif}}\ and\ \bibinfo {author} {\bibfnamefont {S.}~\bibnamefont {Sadiq}},\
  }\href {\doibase 10.1140/epjc/s10052-018-5894-x} {\bibfield  {journal}
  {\bibinfo  {journal} {Eur. Phys. J.}\ }\textbf {\bibinfo {volume} {C78}},\
  \bibinfo {pages} {410} (\bibinfo {year} {2018}{\natexlab{a}})},\ \Eprint
  {http://arxiv.org/abs/1804.09616} {arXiv:1804.09616 [gr-qc]} \BibitemShut
  {NoStop}%
\bibitem [{\citenamefont {Morales}\ and\ \citenamefont
  {Tello-Ortiz}(2018{\natexlab{a}})}]{Morales:2018nmq}%
  \BibitemOpen
  \bibfield  {author} {\bibinfo {author} {\bibfnamefont {E.}~\bibnamefont
  {Morales}}\ and\ \bibinfo {author} {\bibfnamefont {F.}~\bibnamefont
  {Tello-Ortiz}},\ }\href {\doibase 10.1140/epjc/s10052-018-6102-8} {\bibfield
  {journal} {\bibinfo  {journal} {Eur. Phys. J.}\ }\textbf {\bibinfo {volume}
  {C78}},\ \bibinfo {pages} {618} (\bibinfo {year} {2018}{\natexlab{a}})},\
  \Eprint {http://arxiv.org/abs/1805.00592} {arXiv:1805.00592 [gr-qc]}
  \BibitemShut {NoStop}%
\bibitem [{\citenamefont {Sharif}\ and\ \citenamefont
  {Sadiq}(2018{\natexlab{b}})}]{Sharif:2018pzr}%
  \BibitemOpen
  \bibfield  {author} {\bibinfo {author} {\bibfnamefont {M.}~\bibnamefont
  {Sharif}}\ and\ \bibinfo {author} {\bibfnamefont {S.}~\bibnamefont {Sadiq}},\
  }\href {\doibase 10.1140/epjp/i2018-12075-1} {\bibfield  {journal} {\bibinfo
  {journal} {Eur. Phys. J. Plus}\ }\textbf {\bibinfo {volume} {133}},\ \bibinfo
  {pages} {245} (\bibinfo {year} {2018}{\natexlab{b}})}\BibitemShut {NoStop}%
\bibitem [{\citenamefont {Morales}\ and\ \citenamefont
  {Tello-Ortiz}(2018{\natexlab{b}})}]{Morales:2018urp}%
  \BibitemOpen
  \bibfield  {author} {\bibinfo {author} {\bibfnamefont {E.}~\bibnamefont
  {Morales}}\ and\ \bibinfo {author} {\bibfnamefont {F.}~\bibnamefont
  {Tello-Ortiz}},\ }\href {\doibase 10.1140/epjc/s10052-018-6319-6} {\bibfield
  {journal} {\bibinfo  {journal} {Eur. Phys. J.}\ }\textbf {\bibinfo {volume}
  {C78}},\ \bibinfo {pages} {841} (\bibinfo {year} {2018}{\natexlab{b}})},\
  \Eprint {http://arxiv.org/abs/1808.01699} {arXiv:1808.01699 [gr-qc]}
  \BibitemShut {NoStop}%
\bibitem [{\citenamefont {Estrada}\ and\ \citenamefont
  {Prado}(2019)}]{Estrada:2018vrl}%
  \BibitemOpen
  \bibfield  {author} {\bibinfo {author} {\bibfnamefont {M.}~\bibnamefont
  {Estrada}}\ and\ \bibinfo {author} {\bibfnamefont {R.}~\bibnamefont
  {Prado}},\ }\href {\doibase 10.1140/epjp/i2019-12555-8} {\bibfield  {journal}
  {\bibinfo  {journal} {Eur. Phys. J. Plus}\ }\textbf {\bibinfo {volume}
  {134}},\ \bibinfo {pages} {168} (\bibinfo {year} {2019})},\ \Eprint
  {http://arxiv.org/abs/1809.03591} {arXiv:1809.03591 [gr-qc]} \BibitemShut
  {NoStop}%
\bibitem [{\citenamefont {Sharif}\ and\ \citenamefont
  {Saba}(2018)}]{Sharif:2018tiz}%
  \BibitemOpen
  \bibfield  {author} {\bibinfo {author} {\bibfnamefont {M.}~\bibnamefont
  {Sharif}}\ and\ \bibinfo {author} {\bibfnamefont {S.}~\bibnamefont {Saba}},\
  }\href {\doibase 10.1140/epjc/s10052-018-6406-8} {\bibfield  {journal}
  {\bibinfo  {journal} {Eur. Phys. J.}\ }\textbf {\bibinfo {volume} {C78}},\
  \bibinfo {pages} {921} (\bibinfo {year} {2018})},\ \Eprint
  {http://arxiv.org/abs/1811.08112} {arXiv:1811.08112 [gr-qc]} \BibitemShut
  {NoStop}%
\bibitem [{\citenamefont {Ovalle}\ \emph
  {et~al.}(2018{\natexlab{c}})\citenamefont {Ovalle}, \citenamefont {Casadio},
  \citenamefont {da~Rocha}, \citenamefont {Sotomayor},\ and\ \citenamefont
  {Stuchlik}}]{Ovalle:2018ans}%
  \BibitemOpen
  \bibfield  {author} {\bibinfo {author} {\bibfnamefont {J.}~\bibnamefont
  {Ovalle}}, \bibinfo {author} {\bibfnamefont {R.}~\bibnamefont {Casadio}},
  \bibinfo {author} {\bibfnamefont {R.}~\bibnamefont {da~Rocha}}, \bibinfo
  {author} {\bibfnamefont {A.}~\bibnamefont {Sotomayor}}, \ and\ \bibinfo
  {author} {\bibfnamefont {Z.}~\bibnamefont {Stuchlik}},\ }\href {\doibase
  10.1209/0295-5075/124/20004} {\bibfield  {journal} {\bibinfo  {journal}
  {EPL}\ }\textbf {\bibinfo {volume} {124}},\ \bibinfo {pages} {20004}
  (\bibinfo {year} {2018}{\natexlab{c}})},\ \Eprint
  {http://arxiv.org/abs/1811.08559} {arXiv:1811.08559 [gr-qc]} \BibitemShut
  {NoStop}%
\bibitem [{\citenamefont {Contreras}(2019)}]{Contreras:2019fbk}%
  \BibitemOpen
  \bibfield  {author} {\bibinfo {author} {\bibfnamefont {E.}~\bibnamefont
  {Contreras}},\ }\href {\doibase 10.1088/1361-6382/ab11e6} {\bibfield
  {journal} {\bibinfo  {journal} {Class. Quant. Grav.}\ }\textbf {\bibinfo
  {volume} {36}},\ \bibinfo {pages} {095004} (\bibinfo {year} {2019})},\
  \Eprint {http://arxiv.org/abs/1901.00231} {arXiv:1901.00231 [gr-qc]}
  \BibitemShut {NoStop}%
\bibitem [{\citenamefont {Maurya}\ and\ \citenamefont
  {Tello-Ortiz}(2019)}]{Maurya:2019wsk}%
  \BibitemOpen
  \bibfield  {author} {\bibinfo {author} {\bibfnamefont {S.~K.}\ \bibnamefont
  {Maurya}}\ and\ \bibinfo {author} {\bibfnamefont {F.}~\bibnamefont
  {Tello-Ortiz}},\ }\href {\doibase 10.1140/epjc/s10052-019-6602-1} {\bibfield
  {journal} {\bibinfo  {journal} {Eur. Phys. J.}\ }\textbf {\bibinfo {volume}
  {C79}},\ \bibinfo {pages} {85} (\bibinfo {year} {2019})}\BibitemShut
  {NoStop}%
\bibitem [{\citenamefont {Contreras}\ \emph {et~al.}(2019)\citenamefont
  {Contreras}, \citenamefont {Rinc\'on},\ and\ \citenamefont
  {Bargue\~no}}]{Contreras:2019iwm}%
  \BibitemOpen
  \bibfield  {author} {\bibinfo {author} {\bibfnamefont {E.}~\bibnamefont
  {Contreras}}, \bibinfo {author} {\bibfnamefont {A.}~\bibnamefont {Rinc\'on}},
  \ and\ \bibinfo {author} {\bibfnamefont {P.}~\bibnamefont {Bargue\~no}},\
  }\href {\doibase 10.1140/epjc/s10052-019-6749-9} {\bibfield  {journal}
  {\bibinfo  {journal} {Eur. Phys. J.}\ }\textbf {\bibinfo {volume} {C79}},\
  \bibinfo {pages} {216} (\bibinfo {year} {2019})},\ \Eprint
  {http://arxiv.org/abs/1902.02033} {arXiv:1902.02033 [gr-qc]} \BibitemShut
  {NoStop}%
\bibitem [{\citenamefont {Contreras}\ and\ \citenamefont
  {Bargueño}(2019)}]{Contreras:2019mhf}%
  \BibitemOpen
  \bibfield  {author} {\bibinfo {author} {\bibfnamefont {E.}~\bibnamefont
  {Contreras}}\ and\ \bibinfo {author} {\bibfnamefont {P.}~\bibnamefont
  {Bargueño}},\ }\href {\doibase 10.1088/1361-6382/ab47e2} {\bibfield
  {journal} {\bibinfo  {journal} {Class. Quant. Grav.}\ }\textbf {\bibinfo
  {volume} {36}},\ \bibinfo {pages} {215009} (\bibinfo {year} {2019})},\
  \Eprint {http://arxiv.org/abs/1902.09495} {arXiv:1902.09495 [gr-qc]}
  \BibitemShut {NoStop}%
\bibitem [{\citenamefont {Gabbanelli}\ \emph {et~al.}(2019)\citenamefont
  {Gabbanelli}, \citenamefont {Ovalle}, \citenamefont {Sotomayor},
  \citenamefont {Stuchlik},\ and\ \citenamefont
  {Casadio}}]{Gabbanelli:2019txr}%
  \BibitemOpen
  \bibfield  {author} {\bibinfo {author} {\bibfnamefont {L.}~\bibnamefont
  {Gabbanelli}}, \bibinfo {author} {\bibfnamefont {J.}~\bibnamefont {Ovalle}},
  \bibinfo {author} {\bibfnamefont {A.}~\bibnamefont {Sotomayor}}, \bibinfo
  {author} {\bibfnamefont {Z.}~\bibnamefont {Stuchlik}}, \ and\ \bibinfo
  {author} {\bibfnamefont {R.}~\bibnamefont {Casadio}},\ }\href {\doibase
  10.1140/epjc/s10052-019-7022-y} {\bibfield  {journal} {\bibinfo  {journal}
  {Eur. Phys. J. C}\ }\textbf {\bibinfo {volume} {79}},\ \bibinfo {pages} {486}
  (\bibinfo {year} {2019})},\ \Eprint {http://arxiv.org/abs/1905.10162}
  {arXiv:1905.10162 [gr-qc]} \BibitemShut {NoStop}%
\bibitem [{\citenamefont {Estrada}(2019)}]{Estrada:2019aeh}%
  \BibitemOpen
  \bibfield  {author} {\bibinfo {author} {\bibfnamefont {M.}~\bibnamefont
  {Estrada}},\ }\href {\doibase 10.1140/epjc/s10052-019-7444-6} {\bibfield
  {journal} {\bibinfo  {journal} {Eur. Phys. J. C}\ }\textbf {\bibinfo {volume}
  {79}},\ \bibinfo {pages} {918} (\bibinfo {year} {2019})},\ \Eprint
  {http://arxiv.org/abs/1905.12129} {arXiv:1905.12129 [gr-qc]} \BibitemShut
  {NoStop}%
\bibitem [{\citenamefont {Ovalle}\ \emph {et~al.}(2019)\citenamefont {Ovalle},
  \citenamefont {Posada},\ and\ \citenamefont {Stuchlík}}]{Ovalle:2019lbs}%
  \BibitemOpen
  \bibfield  {author} {\bibinfo {author} {\bibfnamefont {J.}~\bibnamefont
  {Ovalle}}, \bibinfo {author} {\bibfnamefont {C.}~\bibnamefont {Posada}}, \
  and\ \bibinfo {author} {\bibfnamefont {Z.}~\bibnamefont {Stuchlík}},\ }\href
  {\doibase 10.1088/1361-6382/ab4461} {\bibfield  {journal} {\bibinfo
  {journal} {Class. Quant. Grav.}\ }\textbf {\bibinfo {volume} {36}},\ \bibinfo
  {pages} {205010} (\bibinfo {year} {2019})},\ \Eprint
  {http://arxiv.org/abs/1905.12452} {arXiv:1905.12452 [gr-qc]} \BibitemShut
  {NoStop}%
\bibitem [{\citenamefont {Maurya}\ and\ \citenamefont
  {Tello-Ortiz}(2020)}]{Maurya:2019hds}%
  \BibitemOpen
  \bibfield  {author} {\bibinfo {author} {\bibfnamefont {S.}~\bibnamefont
  {Maurya}}\ and\ \bibinfo {author} {\bibfnamefont {F.}~\bibnamefont
  {Tello-Ortiz}},\ }\href {\doibase 10.1016/j.dark.2019.100442} {\bibfield
  {journal} {\bibinfo  {journal} {Phys. Dark Univ.}\ }\textbf {\bibinfo
  {volume} {27}},\ \bibinfo {pages} {100442} (\bibinfo {year} {2020})},\
  \Eprint {http://arxiv.org/abs/1905.13519} {arXiv:1905.13519 [gr-qc]}
  \BibitemShut {NoStop}%
\bibitem [{\citenamefont {Hensh}\ and\ \citenamefont
  {Stuchlík}(2019)}]{Hensh:2019rtb}%
  \BibitemOpen
  \bibfield  {author} {\bibinfo {author} {\bibfnamefont {S.}~\bibnamefont
  {Hensh}}\ and\ \bibinfo {author} {\bibfnamefont {Z.~e.}\ \bibnamefont
  {Stuchlík}},\ }\href {\doibase 10.1140/epjc/s10052-019-7360-9} {\bibfield
  {journal} {\bibinfo  {journal} {Eur. Phys. J. C}\ }\textbf {\bibinfo {volume}
  {79}},\ \bibinfo {pages} {834} (\bibinfo {year} {2019})},\ \Eprint
  {http://arxiv.org/abs/1906.08368} {arXiv:1906.08368 [gr-qc]} \BibitemShut
  {NoStop}%
\bibitem [{\citenamefont {Linares~Cedeño}\ and\ \citenamefont
  {Contreras}(2020)}]{Cedeno:2019qkf}%
  \BibitemOpen
  \bibfield  {author} {\bibinfo {author} {\bibfnamefont {F.~X.}\ \bibnamefont
  {Linares~Cedeño}}\ and\ \bibinfo {author} {\bibfnamefont {E.}~\bibnamefont
  {Contreras}},\ }\href {\doibase 10.1016/j.dark.2020.100543} {\bibfield
  {journal} {\bibinfo  {journal} {Phys. Dark Univ.}\ }\textbf {\bibinfo
  {volume} {28}},\ \bibinfo {pages} {100543} (\bibinfo {year} {2020})},\
  \Eprint {http://arxiv.org/abs/1907.04892} {arXiv:1907.04892 [gr-qc]}
  \BibitemShut {NoStop}%
\bibitem [{\citenamefont {León}\ and\ \citenamefont
  {Sotomayor}(2019)}]{Leon:2019abq}%
  \BibitemOpen
  \bibfield  {author} {\bibinfo {author} {\bibfnamefont {P.}~\bibnamefont
  {León}}\ and\ \bibinfo {author} {\bibfnamefont {A.}~\bibnamefont
  {Sotomayor}},\ }\href {\doibase 10.1002/prop.201900077} {\bibfield  {journal}
  {\bibinfo  {journal} {Fortsch. Phys.}\ }\textbf {\bibinfo {volume} {67}},\
  \bibinfo {pages} {1900077} (\bibinfo {year} {2019})},\ \Eprint
  {http://arxiv.org/abs/1907.11763} {arXiv:1907.11763 [gr-qc]} \BibitemShut
  {NoStop}%
\bibitem [{\citenamefont {Torres-Sánchez}\ and\ \citenamefont
  {Contreras}(2019)}]{Torres:2019mee}%
  \BibitemOpen
  \bibfield  {author} {\bibinfo {author} {\bibfnamefont {V.}~\bibnamefont
  {Torres-Sánchez}}\ and\ \bibinfo {author} {\bibfnamefont {E.}~\bibnamefont
  {Contreras}},\ }\href {\doibase 10.1140/epjc/s10052-019-7341-z} {\bibfield
  {journal} {\bibinfo  {journal} {Eur. Phys. J. C}\ }\textbf {\bibinfo {volume}
  {79}},\ \bibinfo {pages} {829} (\bibinfo {year} {2019})},\ \Eprint
  {http://arxiv.org/abs/1908.08194} {arXiv:1908.08194 [gr-qc]} \BibitemShut
  {NoStop}%
\bibitem [{\citenamefont {Casadio}\ \emph {et~al.}(2019)\citenamefont
  {Casadio}, \citenamefont {Contreras}, \citenamefont {Ovalle}, \citenamefont
  {Sotomayor},\ and\ \citenamefont {Stuchlick}}]{Casadio:2019usg}%
  \BibitemOpen
  \bibfield  {author} {\bibinfo {author} {\bibfnamefont {R.}~\bibnamefont
  {Casadio}}, \bibinfo {author} {\bibfnamefont {E.}~\bibnamefont {Contreras}},
  \bibinfo {author} {\bibfnamefont {J.}~\bibnamefont {Ovalle}}, \bibinfo
  {author} {\bibfnamefont {A.}~\bibnamefont {Sotomayor}}, \ and\ \bibinfo
  {author} {\bibfnamefont {Z.}~\bibnamefont {Stuchlick}},\ }\href {\doibase
  10.1140/epjc/s10052-019-7358-3} {\bibfield  {journal} {\bibinfo  {journal}
  {Eur. Phys. J. C}\ }\textbf {\bibinfo {volume} {79}},\ \bibinfo {pages} {826}
  (\bibinfo {year} {2019})},\ \Eprint {http://arxiv.org/abs/1909.01902}
  {arXiv:1909.01902 [gr-qc]} \BibitemShut {NoStop}%
\bibitem [{\citenamefont {Singh}\ \emph {et~al.}(2019)\citenamefont {Singh},
  \citenamefont {Maurya}, \citenamefont {Jasim},\ and\ \citenamefont
  {Rahaman}}]{Singh:2019ktp}%
  \BibitemOpen
  \bibfield  {author} {\bibinfo {author} {\bibfnamefont {K.}~\bibnamefont
  {Singh}}, \bibinfo {author} {\bibfnamefont {S.}~\bibnamefont {Maurya}},
  \bibinfo {author} {\bibfnamefont {M.}~\bibnamefont {Jasim}}, \ and\ \bibinfo
  {author} {\bibfnamefont {F.}~\bibnamefont {Rahaman}},\ }\href {\doibase
  10.1140/epjc/s10052-019-7377-0} {\bibfield  {journal} {\bibinfo  {journal}
  {Eur. Phys. J. C}\ }\textbf {\bibinfo {volume} {79}},\ \bibinfo {pages} {851}
  (\bibinfo {year} {2019})}\BibitemShut {NoStop}%
\bibitem [{\citenamefont {Maurya}(2019)}]{Maurya:2019noq}%
  \BibitemOpen
  \bibfield  {author} {\bibinfo {author} {\bibfnamefont {S.}~\bibnamefont
  {Maurya}},\ }\href {\doibase 10.1140/epjc/s10052-019-7458-0} {\bibfield
  {journal} {\bibinfo  {journal} {Eur. Phys. J. C}\ }\textbf {\bibinfo {volume}
  {79}},\ \bibinfo {pages} {958} (\bibinfo {year} {2019})}\BibitemShut
  {NoStop}%
\bibitem [{\citenamefont {Sharif}\ and\ \citenamefont
  {Waseem}(2019)}]{Sharif:2019mjn}%
  \BibitemOpen
  \bibfield  {author} {\bibinfo {author} {\bibfnamefont {M.}~\bibnamefont
  {Sharif}}\ and\ \bibinfo {author} {\bibfnamefont {A.}~\bibnamefont
  {Waseem}},\ }\href {\doibase 10.1016/j.aop.2019.03.003} {\bibfield  {journal}
  {\bibinfo  {journal} {Annals Phys.}\ }\textbf {\bibinfo {volume} {405}},\
  \bibinfo {pages} {14} (\bibinfo {year} {2019})}\BibitemShut {NoStop}%
\bibitem [{\citenamefont {Abellán}\ \emph {et~al.}(2020)\citenamefont
  {Abellán}, \citenamefont {Torres-Sánchez}, \citenamefont {Fuenmayor},\ and\
  \citenamefont {Contreras}}]{Abellan:2020wjw}%
  \BibitemOpen
  \bibfield  {author} {\bibinfo {author} {\bibfnamefont {G.}~\bibnamefont
  {Abellán}}, \bibinfo {author} {\bibfnamefont {V.}~\bibnamefont
  {Torres-Sánchez}}, \bibinfo {author} {\bibfnamefont {E.}~\bibnamefont
  {Fuenmayor}}, \ and\ \bibinfo {author} {\bibfnamefont {E.}~\bibnamefont
  {Contreras}},\ }\href {\doibase 10.1140/epjc/s10052-020-7749-5} {\bibfield
  {journal} {\bibinfo  {journal} {Eur. Phys. J. C}\ }\textbf {\bibinfo {volume}
  {80}},\ \bibinfo {pages} {177} (\bibinfo {year} {2020})},\ \Eprint
  {http://arxiv.org/abs/2001.08573} {arXiv:2001.08573 [gr-qc]} \BibitemShut
  {NoStop}%
\bibitem [{\citenamefont {Sharif}\ and\ \citenamefont
  {Ama-Tul-Mughani}(2020)}]{Sharif:2020vvk}%
  \BibitemOpen
  \bibfield  {author} {\bibinfo {author} {\bibfnamefont {M.}~\bibnamefont
  {Sharif}}\ and\ \bibinfo {author} {\bibfnamefont {Q.}~\bibnamefont
  {Ama-Tul-Mughani}},\ }\href {\doibase 10.1016/j.aop.2020.168122} {\bibfield
  {journal} {\bibinfo  {journal} {Annals Phys.}\ }\textbf {\bibinfo {volume}
  {415}},\ \bibinfo {pages} {168122} (\bibinfo {year} {2020})},\ \Eprint
  {http://arxiv.org/abs/2004.07925} {arXiv:2004.07925 [gr-qc]} \BibitemShut
  {NoStop}%
\bibitem [{\citenamefont {Tello-Ortiz}(2020)}]{Tello-Ortiz:2020ydf}%
  \BibitemOpen
  \bibfield  {author} {\bibinfo {author} {\bibfnamefont {F.}~\bibnamefont
  {Tello-Ortiz}},\ }\href {\doibase 10.1140/epjc/s10052-020-7995-6} {\bibfield
  {journal} {\bibinfo  {journal} {Eur. Phys. J. C}\ }\textbf {\bibinfo {volume}
  {80}},\ \bibinfo {pages} {413} (\bibinfo {year} {2020})}\BibitemShut
  {NoStop}%
\bibitem [{\citenamefont {Maurya}(2020)}]{Maurya:2020rny}%
  \BibitemOpen
  \bibfield  {author} {\bibinfo {author} {\bibfnamefont {S.}~\bibnamefont
  {Maurya}},\ }\href {\doibase 10.1140/epjc/s10052-020-7993-8} {\bibfield
  {journal} {\bibinfo  {journal} {Eur. Phys. J. C}\ }\textbf {\bibinfo {volume}
  {80}},\ \bibinfo {pages} {429} (\bibinfo {year} {2020})}\BibitemShut
  {NoStop}%
\bibitem [{\citenamefont {Rinc\'on}\ \emph {et~al.}(2020)\citenamefont
  {Rinc\'on}, \citenamefont {Contreras}, \citenamefont {Tello-Ortiz},
  \citenamefont {Bargueño},\ and\ \citenamefont {Abell\'an}}]{Rincon:2020izv}%
  \BibitemOpen
  \bibfield  {author} {\bibinfo {author} {\bibfnamefont {A.}~\bibnamefont
  {Rinc\'on}}, \bibinfo {author} {\bibfnamefont {E.}~\bibnamefont {Contreras}},
  \bibinfo {author} {\bibfnamefont {F.}~\bibnamefont {Tello-Ortiz}}, \bibinfo
  {author} {\bibfnamefont {P.}~\bibnamefont {Bargueño}}, \ and\ \bibinfo
  {author} {\bibfnamefont {G.}~\bibnamefont {Abell\'an}},\ }\href {\doibase
  10.1140/epjc/s10052-020-8071-y} {\bibfield  {journal} {\bibinfo  {journal}
  {Eur. Phys. J. C}\ }\textbf {\bibinfo {volume} {80}},\ \bibinfo {pages} {490}
  (\bibinfo {year} {2020})},\ \Eprint {http://arxiv.org/abs/2005.10991}
  {arXiv:2005.10991 [gr-qc]} \BibitemShut {NoStop}%
\bibitem [{\citenamefont {Sharif}\ and\ \citenamefont
  {Majid}(2020)}]{Sharif:2020arn}%
  \BibitemOpen
  \bibfield  {author} {\bibinfo {author} {\bibfnamefont {M.}~\bibnamefont
  {Sharif}}\ and\ \bibinfo {author} {\bibfnamefont {A.}~\bibnamefont {Majid}},\
  }\href {\doibase 10.1016/j.dark.2020.100610} {\bibfield  {journal} {\bibinfo
  {journal} {Phys. Dark Univ.}\ }\textbf {\bibinfo {volume} {30}},\ \bibinfo
  {pages} {100610} (\bibinfo {year} {2020})},\ \Eprint
  {http://arxiv.org/abs/2006.04578} {arXiv:2006.04578 [gr-qc]} \BibitemShut
  {NoStop}%
\bibitem [{\citenamefont {Maurya}\ \emph {et~al.}(2020)\citenamefont {Maurya},
  \citenamefont {Singh},\ and\ \citenamefont {Dayanandan}}]{Maurya:2020gjw}%
  \BibitemOpen
  \bibfield  {author} {\bibinfo {author} {\bibfnamefont {S.}~\bibnamefont
  {Maurya}}, \bibinfo {author} {\bibfnamefont {K.~N.}\ \bibnamefont {Singh}}, \
  and\ \bibinfo {author} {\bibfnamefont {B.}~\bibnamefont {Dayanandan}},\
  }\href {\doibase 10.1140/epjc/s10052-020-8005-8} {\bibfield  {journal}
  {\bibinfo  {journal} {Eur. Phys. J. C}\ }\textbf {\bibinfo {volume} {80}},\
  \bibinfo {pages} {448} (\bibinfo {year} {2020})}\BibitemShut {NoStop}%
\bibitem [{\citenamefont {Herrera}\ and\ \citenamefont
  {Santos}(1997)}]{Herrera:1997plx}%
  \BibitemOpen
  \bibfield  {author} {\bibinfo {author} {\bibfnamefont {L.}~\bibnamefont
  {Herrera}}\ and\ \bibinfo {author} {\bibfnamefont {N.~O.}\ \bibnamefont
  {Santos}},\ }\href {\doibase 10.1016/S0370-1573(96)00042-7} {\bibfield
  {journal} {\bibinfo  {journal} {Phys. Rept.}\ }\textbf {\bibinfo {volume}
  {286}},\ \bibinfo {pages} {53} (\bibinfo {year} {1997})}\BibitemShut
  {NoStop}%
\bibitem [{\citenamefont {Mak}\ and\ \citenamefont {Harko}(2003)}]{Mak:2001eb}%
  \BibitemOpen
  \bibfield  {author} {\bibinfo {author} {\bibfnamefont {M.~K.}\ \bibnamefont
  {Mak}}\ and\ \bibinfo {author} {\bibfnamefont {T.}~\bibnamefont {Harko}},\
  }\href {\doibase 10.1098/rspa.2002.1014} {\bibfield  {journal} {\bibinfo
  {journal} {Proc. Roy. Soc. Lond.}\ }\textbf {\bibinfo {volume} {A459}},\
  \bibinfo {pages} {393} (\bibinfo {year} {2003})},\ \Eprint
  {http://arxiv.org/abs/gr-qc/0110103} {arXiv:gr-qc/0110103 [gr-qc]}
  \BibitemShut {NoStop}%
\bibitem [{\citenamefont {Kiselev}(2003)}]{Kiselev:2002dx}%
  \BibitemOpen
  \bibfield  {author} {\bibinfo {author} {\bibfnamefont {V.}~\bibnamefont
  {Kiselev}},\ }\href {\doibase 10.1088/0264-9381/20/6/310} {\bibfield
  {journal} {\bibinfo  {journal} {Class. Quant. Grav.}\ }\textbf {\bibinfo
  {volume} {20}},\ \bibinfo {pages} {1187} (\bibinfo {year} {2003})},\ \Eprint
  {http://arxiv.org/abs/gr-qc/0210040} {arXiv:gr-qc/0210040} \BibitemShut
  {NoStop}%
\bibitem [{\citenamefont {Visser}(2020)}]{Visser:2019brz}%
  \BibitemOpen
  \bibfield  {author} {\bibinfo {author} {\bibfnamefont {M.}~\bibnamefont
  {Visser}},\ }\href {\doibase 10.1088/1361-6382/ab60b8} {\bibfield  {journal}
  {\bibinfo  {journal} {Class. Quant. Grav.}\ }\textbf {\bibinfo {volume}
  {37}},\ \bibinfo {pages} {045001} (\bibinfo {year} {2020})},\ \Eprint
  {http://arxiv.org/abs/1908.11058} {arXiv:1908.11058 [gr-qc]} \BibitemShut
  {NoStop}%
\bibitem [{\citenamefont {Visser}(1995)}]{Visser:1995cc}%
  \BibitemOpen
  \bibfield  {author} {\bibinfo {author} {\bibfnamefont {M.}~\bibnamefont
  {Visser}},\ }\href@noop {} {\emph {\bibinfo {title} {{Lorentzian wormholes:
  From Einstein to Hawking}}}}\ (\bibinfo {year} {1995})\BibitemShut {NoStop}%
\bibitem [{\citenamefont {Curiel}(2017)}]{Curiel:2014zba}%
  \BibitemOpen
  \bibfield  {author} {\bibinfo {author} {\bibfnamefont {E.}~\bibnamefont
  {Curiel}},\ }\enquote {\bibinfo {title} {{A Primer on Energy Conditions}},}\
  \ (\bibinfo {year} {2017})\ pp.\ \bibinfo {pages} {43--104},\ \Eprint
  {http://arxiv.org/abs/1405.0403} {arXiv:1405.0403 [physics.hist-ph]}
  \BibitemShut {NoStop}%
\bibitem [{\citenamefont {Salazar}\ \emph {et~al.}(1987)\citenamefont
  {Salazar}, \citenamefont {Garcia},\ and\ \citenamefont
  {Plebanski}}]{Salazar:1987ap}%
  \BibitemOpen
  \bibfield  {author} {\bibinfo {author} {\bibfnamefont {I.}~\bibnamefont
  {Salazar}}, \bibinfo {author} {\bibfnamefont {A.}~\bibnamefont {Garcia}}, \
  and\ \bibinfo {author} {\bibfnamefont {J.}~\bibnamefont {Plebanski}},\ }\href
  {\doibase 10.1063/1.527430} {\bibfield  {journal} {\bibinfo  {journal} {J.
  Math. Phys.}\ }\textbf {\bibinfo {volume} {28}},\ \bibinfo {pages} {2171}
  (\bibinfo {year} {1987})}\BibitemShut {NoStop}%
\bibitem [{\citenamefont {Ayon-Beato}\ and\ \citenamefont
  {Garcia}(1998)}]{AyonBeato:1998ub}%
  \BibitemOpen
  \bibfield  {author} {\bibinfo {author} {\bibfnamefont {E.}~\bibnamefont
  {Ayon-Beato}}\ and\ \bibinfo {author} {\bibfnamefont {A.}~\bibnamefont
  {Garcia}},\ }\href {\doibase 10.1103/PhysRevLett.80.5056} {\bibfield
  {journal} {\bibinfo  {journal} {Phys. Rev. Lett.}\ }\textbf {\bibinfo
  {volume} {80}},\ \bibinfo {pages} {5056} (\bibinfo {year} {1998})},\ \Eprint
  {http://arxiv.org/abs/gr-qc/9911046} {arXiv:gr-qc/9911046} \BibitemShut
  {NoStop}%
\end{thebibliography}%
\bibliographystyle{apsrev4-1.bst}
%
%
\end{document}